\documentclass[%
 aip,
 amsmath,amssymb,
preprint,%
]{revtex4-2}

\usepackage{graphicx}
\usepackage{dcolumn}
\usepackage{bm}

\usepackage[utf8]{inputenc}
\usepackage[T1]{fontenc}

\usepackage{etoolbox}

\usepackage{xcolor}

\makeatletter
\def\@email#1#2{%
 \endgroup
 \patchcmd{\titleblock@produce}
  {\frontmatter@RRAPformat}
  {\frontmatter@RRAPformat{\produce@RRAP{*#1\href{mailto:#2}{#2}}}\frontmatter@RRAPformat}
  {}{}
}%
\usepackage{setspace}
\makeatother
\newcommand{\markchange}[1]{{\color{black}#1}}
\begin{document}
\title{Coherent Biexciton Transport in the Presence of Exciton–Exciton Annihilation in Molecular Aggregates}

\author{Rajesh Dutta}
\author{Chern Chuang}

\affiliation{University of Nevada, Las Vegas, Department of Chemistry and Biochemistry,\\ 4505 S Maryland Pkwy, NV 89154}

\begin{abstract}
Biexciton dynamics in molecular aggregates provides a sensitive probe of the interplay between quantum coherence, band structure, and dissipation under strong excitation conditions. We present a theoretical framework for biexciton dynamics in molecular aggregates that explicitly treats populations and coherences across excitation manifolds within a reduced density-matrix formalism. By extending kinetic descriptions beyond the weak-coupling limit, the approach captures the influence of exciton delocalization and exciton-exciton annihilation while remaining computationally tractable within a Markovian description of environmental relaxation. Using this framework, we investigate how the spatial profile and momentum composition of the initial biexciton state govern fluorescence decay and transport. Incoherent initial conditions lead to strongly non-exponential relaxation and time-dependent diffusion driven by nonlinear population kinetics. In contrast, coherently prepared biexciton states exhibit pronounced early-time coherent transport, whose character depends sensitively on whether the initial state is prepared as a standing-wave or traveling-wave superposition of single-exciton modes. Despite nearly identical emission dynamics for J and H aggregate, biexciton transport properties differ markedly due to band structure-dependent interference effect. Our results demonstrate that biexciton dynamics remains strongly influenced by initial-state coherence and momentum composition. Besides initial-state preparation, the coherent-to-incoherent crossover and the diffusive spreading of the exciton density are sensitive to internal conversion processes such as exciton fusion and the decay to the first excited state. The present work establishes initial-state preparation as a key control parameter for many-exciton transport in excitonic systems and provides a general framework for interpreting nonlinear optical experiments beyond population-based descriptions.
\end{abstract}

\maketitle

\section{Introduction}
Exciton transport in natural light-harvesting complexes, artificial molecular aggregates, and conjugated polymers has attracted substantial interest over decades, particularly following the recognition of the role of quantum coherence in these systems. \cite{engel2007evidence,calhoun2009quantum,collini2009coherent,collini2010coherently}Both experimental\cite{engel2007evidence,calhoun2009quantum,collini2010coherently,collini2009coherent,collini2009coherent,maly2020wavelike} and theoretical\cite{ishizaki2009theoretical,ishizaki2009unified,moix2011efficient,barford2014theory,bittner2014noise,kassal2013does,dutta2019delocalization} advances have sought to unravel the mechanisms underlying efficient energy transport, with potential applications in renewable energy technologies such as photovoltaics and solar cells.\cite{cao2009optimization,wu2010efficient,chen2011excitation,zhang2015delocalized,zhang2016dark,sarkar2020environment,dutta2021excitation}

Most theoretical studies have focused on quantum transport in the single-exciton regime, with the primary aim of elucidating the interplay between static disorder and environmental noise in molecular and materials of different dimensionalities and geometries using dynamical and optical properties.\cite{haken1972coupled,haken1973exactly,silbey1976electronic,moix2013coherent,yuen2014coherent,lee2015coherent,dutta2016effects,chuang2016quantum,chuang2021universal,karamlou2022quantum,jansen2024electronically,blach2025environment,barford2024using,dutta2025quantum} \markchange{In one- and two-dimensional systems, any finite amount of disorder leads to the suppression of transport. While moderate environmental noise can enhance transport by suppressing localization effects (often referred to as environment-assisted transport),\cite{rebentrost2009environment,moix2013coherent} this behavior is not universal. In particular, when the thermal energy ($k_B T$) or the corresponding dephasing strength exceeds the exciton bandwidth, strong noise leads to rapid decoherence and suppresses coherent motion, resulting in a reduction of the diffusion coefficient.\cite{moix2013coherent,barford2024using} In this high-temperature or strong-dephasing (``Quantum Zeno'') regime, transport becomes increasingly classical and can scale inversely with temperature,\cite{barford2024using} reflecting noise-induced localization of dynamics.} However, the single excitonic description is valid when at most one molecule is in the excited state while the remaining molecules are in their ground states, a condition that is naturally satisfied under a weak field. Such weak-excitation conditions are naturally realized under solar irradiation and are relevant for both biological photosynthesis and photovoltaic systems. 

On the other hand, in laser-based experiments the single-exciton approximation can break down, since at high excitation densities or under strong laser fluence multiple excitons may be generated in proximity. The formation of multiexciton states gives rise to exciton–exciton interactions, which can be directly probed through nonlinear optical responses.\cite{spano1989nonlinear,kuhn1996two} In a seminal study, the nonlinear absorption coefficient of a one-dimensional molecular aggregate composed of coupled two-level systems in the presence of site disorder was evaluated.\cite{spano1991fermion} Experimentally, the first observation of a one to two-exciton transition in J aggregates was reported by Fidder \textit{et. al.}.\cite{fidder1993observation} Subsequently, an analytic expression for the exciton–exciton scattering amplitude in an infinite, periodic one-dimensional aggregate was derived, providing a theoretical foundation for interpreting the characteristic spectral line shapes of the dominant J-band resonance observed in coherent third-order nonlinear spectroscopy.\cite{abramavicius2013exciton} 

In the presence of multiexciton states, exciton–exciton annihilation (EEA) becomes an essential, if not dominant, component of the dynamics.\cite{valkunas2009exciton} This process is commonly described as a two-step mechanism: first, two excitations interact, resulting in the formation of a higher excited state on one molecule while the other molecule relaxes to the ground state, a process often referred to as exciton fusion. In the second step, rapid internal conversion relaxes the higher excited state to the first excited state. 

Experimentally, EEA can be characterized using third-order nonlinear spectroscopy as a function of laser fluence,\cite{bittner1994ultrafast,barzda2001singlet,dostal2018direct,maly2018signatures,kumar2023exciton,sohoni2024optically,bubilaitis2024signatures} and such EEA signal is commonly modeled as an incoherent, classical reaction governed by a bimolecular rate law.\cite{sundstrom1988annihilation,van1995dynamics,gadonas1997wavelength,van2000photosynthetic,valkunas1995nonlinear,Wan2015,Deng2020} However, when excitons are strongly delocalized and/or at low temperatures, this phenomenological description is no longer adequate. Beyond phenomenological rate-based descriptions, several microscopic theoretical frameworks for EEA have been developed.\cite{gaivzauskas1995annihilation,malyshev1999exciton,ryzhov2001low} In particular, density-matrix–based approaches have been formulated to capture the role of quantum coherence and exciton delocalization.\cite{renger1997theory,renger1997multiple,bruggemann2001microscopic} More recent studies have employed Fermi’s golden rule to describe EEA in molecular aggregates, explicitly accounting for the sign of the dipole–dipole coupling that distinguishes J- and H-type aggregates.\cite{tempelaar2017exciton} Both theoretical analyses and experimental observations have demonstrated that, in H-aggregates, coherent suppression of EEA arises from the destructive interference originating from their out-of-phase excitonic wavefunctions.\cite{kumar2023exciton,sohoni2024optically} Furthermore, by systematically reducing the density-matrix description, effective kinetic equations involving multiple excitation manifolds were derived and successfully applied to describe annihilation dynamics in conjugated polymer systems.\cite{may2014kinetic,hader2016identification} 

Despite these advances, a comprehensive theoretical framework that explicitly accounts for coherences within and between different excitation manifolds such as single-exciton, multiexciton, and their cross-manifold coherences—and their impact on the spatiotemporal dynamics of molecular aggregates is still missing. In this work, we pursue two primary objectives. First, we extend the kinetic formulation originally proposed by May\cite{may2014kinetic} to regimes of intermediate and strong coupling, where coherent effects are expected to play a significant role and cannot be neglected. Second, while coherent and incoherent control at the single-exciton level—both temporal and spatial,\cite{hoki2011excitation,jankovic2020exact,yang2020steady,jung2020energy,dutta2024memory,tutunnikov2023coherent} has been extensively explored in light-harvesting systems and molecular aggregates, and temporal control of multiexciton processes\cite{may2014kinetic,wang2018laser} has been addressed theoretically, a corresponding framework for spatial profile dependence and for assessing the role of initial state preparation in multiexcitonic dynamics is lacking. Here, we develop such a theory and apply it to biexciton dynamics in both J- and H-aggregates. Importantly, while we focus here on the homogeneous limit where localization is not a factor, the present approach is general regardless of localization landscape and can be systematically extended to higher multiexciton manifolds.

The remainder of this paper is organized as follows. In Sec. II, we introduce the model and Hamiltonian. Section III derives the equations of motion for populations and coherences in a multiexcitonic framework. In Sec. IV, we construct biexciton initial states from which full quantum dynamics are followed. Section V examines the implications of biexciton dynamics across different regimes, with emphasis on experimentally relevant observables such as time-resolved emission and exciton diffusion. Finally, Sec. VI summarizes the paper.
\section{Model and Hamiltonian}
Under weak optical excitation, the electronic Hilbert space can be restricted to the overall ground state and states containing at most a single excitation. In this regime, the dynamics is conveniently described in terms of localized molecular excitations that can delocalize via intermolecular Coulomb interactions. In contrast, under strong excitation conditions, EEA becomes significant and suppresses the formation of long-lived multiexciton states. To describe biexciton dynamics in the presence of EEA in a linear molecular  with three level systems,\cite{knoester1995unusual} we adopt the Hamiltonian proposed by May\cite{may2014kinetic}
\begin{equation}
H_{\mathrm{tot}}=H^{(e)}+H^{(f)}+V^{(ef)}
\end{equation}
where $H^{(e)}$ accounts for the singly excited manyfold, $H^{(f)}$ denotes the third-level excitation manifold, and the cross term $V^{(ef)}$ represents the fusion/fission coupling between the two manifolds and is ultimately responsible for EEA. 

The first excited state Frenkel Hamiltonian can be written as
\begin{equation}
H^{(e)} = \sum_{m} E_m B_m^{\dagger} B_m 
+ \sum_{m\neq n} J_{mn} B_m^{\dagger} B_n 
\end{equation}

Similarly, second excited state Hamiltonian can be expressed as
\begin{equation}
H^{(f)} = \sum_{m} \varepsilon_m D_m^{\dagger} D_m + \sum_{m\neq n} \mathcal{J}_{mn} D_m^{\dagger} D_n
\end{equation}

\begin{figure*}
 \centering
 \includegraphics[height=4cm]{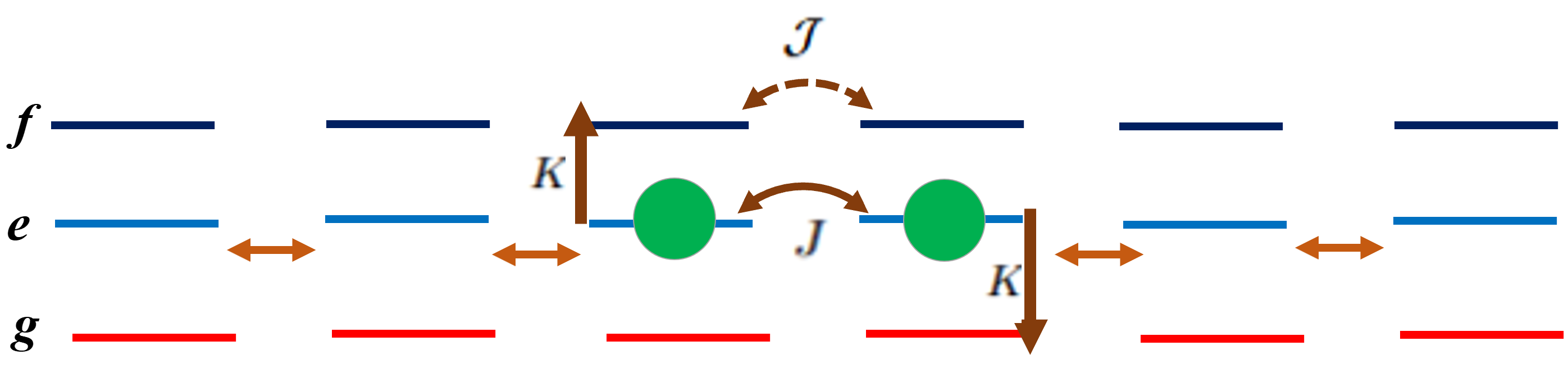}
\caption{Biexciton dynamics in linear chain of three-level systems with nearest neighbor interaction. $J$ is the interaction between first excited state, $\mathcal{J}$ refers to second excited state interaction and $K$ denotes fusion due to exciton-exciton annhiliation.}
 \label{fig:fig1}
\end{figure*}

The interaction term can be given as
\begin{equation}
V^{(ef)}=\sum_{m\neq n}
\left( K_{mn} D_m^{\dagger} B_n + \text{H.c.} \right)
\end{equation}
 Here, $J_{mn}$ denotes the coupling between the first excited states ($e$) of different molecules, while $\mathcal{J}_{mn}$ represents the interaction between the second (higher) excited states ($f$) of different molecules. To account for EEA, the energy difference between the ground and first excited states, $E_m$, is assumed to be comparable to the energy separation between the first and second excited states, $\epsilon_m$. Under this resonance condition, biexciton states can merge by promoting one molecule to the second excited state while simultaneously relaxing the other molecule to the ground state. The coupling $K_{mn}$ accounts for the formation of second excited states by fusion of first excited states (Fig.~\ref{fig:fig1}). $B_m^{\dagger}$ and $B_m$ ($D_m^{\dagger}$ and $D_m$) represent transition operators between zero- (one-) and one- (two-) excitation manifolds. We also assume that exciton-exciton interactions can be neglected but the general framework remains the same if those were accounted for.

We consider a homogeneous chain of molecules where all the couplings are nearest neighbor and coefficients are real i.e.

\begin{equation}
\begin{split}
& \varepsilon_m = 2E_m = 2E, \\ 
& J_{mn} = J \left( \delta_{m,n+1} + \delta_{m,n-1} \right),  \\
& \mathcal{J}_{mn} = \mathcal{J} \left( \delta_{m,n+1} + \delta_{m,n-1} \right),  \\
& K_{mn} = K \left( \delta_{m,n+1} + \delta_{m,n-1} \right)
\end{split}
\end{equation}
\markchange{ The parameter $J$ represents the standard resonant excitonic coupling between molecular sites, arising from Coulomb interactions between transition charge densities, and governs single-exciton transport between first excited states ($e$). In contrast, $\mathcal{J}$ and $K$ describe couplings in the multi-exciton manifold. When two molecules are simultaneously excited to the first excited state, their interaction can lead to exciton–exciton annihilation, where one excitation is promoted to a higher excited state ($f$) while the other relaxes to the ground state ($g$), as shown in Fig.~\ref{fig:fig1}. The coupling $\mathcal{J}$ accounts for transitions involving higher excited states, while $K$ characterizes the effective interaction responsible for exciton fusion/annihilation processes. All these couplings originate from intermolecular Coulomb interactions; however, $J$ ($\mathcal{J}$) describes one-exciton transfer within the first (second) excited state manifold, whereas $K$ captures many-body processes involving multiple excitations.} Again our framework is not restricted to homogeneous systems and analysis for disordered systems will be reserved for future work. 

\section{Equation of motion}
In the presence of a dissipative environment, a Markovian master equation for such a system can be expressed in the Lindblad form as follows
\begin{equation}
\label{eq:lindblad}
\frac{\partial}{\partial t}\operatorname{tr}\{\hat{\rho}(t)\,\hat{O}\}
= \frac{i}{\hbar}\left\langle[\hat{H},\hat{O}]_{-}\right\rangle
- \left\langle\tilde{D}\,\hat{O}\right\rangle \,.
\end{equation}
Here, an arbitrary operator $\hat{O}$, constructed from different combinations of transition operators, is introduced. This procedure effectively projects out the relevant diagonal and off-diagonal reduced density-matrix elements, thereby reducing the computational cost associated with solving the full density-matrix dynamics. 
\begin{equation}
\label{eq:dissipator_1}
\mathcal{\tilde{D}}_{e} \hat{O}
= \frac{\kappa}{2}\sum_{m}  \left( [B_{m}^{\dagger} B_{m}, \hat{O}]_{+} - 2 B_{m}^{\dagger} \hat{O} B_{m} \right)
\end{equation}

\begin{equation}
\label{eq:dissipator_2}
\mathcal{\tilde{D}}_{f} \hat{O}
= \frac{r}{2}\sum_{n}  \left( [D_{n}^{\dagger} D_{n}, \hat{O}]_{+} - 2 D_{n}^{\dagger} \hat{O} D_{n} \right)
\end{equation}
Here, $\kappa$ represents the decay rate from the first excited to ground state and $r$ denotes the decay from second to first excited state. 

{\markchange{The validity of the Markovian approximation can be understood from the separation of characteristic timescales. The lifetime of the first excited state ($\kappa^{-1}$) is on the order of nanoseconds, while exciton transport driven by intersite coupling ($J$) occurs on picosecond timescales. In contrast, exciton–exciton annihilation involves ultrafast processes, including fusion ($K^{-1}$) and subsequent relaxation from the higher excited state ($r^{-1}$), which typically occur on femtosecond timescales.\cite{hader2016identification} This separation of timescales implies that the inclusion of memory-induced system–bath correlations could modify the transient behavior of the dynamics, while the overall transport and emission dynamics at longer times remain well captured within the present framework.}}

Population of the ground, first and second excited states, respectively, are given by

\begin{equation}
\begin{split}
\label{eq:pop_states}
& \Pi_n = \langle B_{n} B_{n}^{\dagger} \rangle, \\ 
& P_n = \langle B_{n}^{\dagger} B_{n} \rangle, \\ 
& N_n = \langle D_{n}^{\dagger} D_{n} \rangle
\end{split}
\end{equation}

The completeness relation can thus be written as 
\begin{equation}
\label{eq:completeness}
\Pi_n+ P_n +N_n=1
\end{equation}
{\markchange{which expresses conservation of probability and ensures that, at any time, each molecule occupies one of the three allowed levels.}}
Equation of motion for the diagonal part reads\cite{may2014kinetic}
\begin{align}
\label{eq:population_eqn}
\frac{\partial}{\partial t} P_m(t)
&= -\kappa P_m + rN_m  \nonumber\\
&\quad +\frac{2}{\hbar}\sum_n J_{mn}\, \text{Im}[W_{mn}]
   -\frac{2}{\hbar}\sum_n \mathcal{J}_{mn}\, \text{Im}[Z_{mn}] \nonumber\\
&\quad -\frac{2}{\hbar}\sum_n K_{mn}\, \text{Im}[R_{mn}]
   -\frac{2}{\hbar}\sum_n K_{nm}\, \text{Im}[R_{nm}]
\end{align}

\begin{equation}
\label{eq:pop_eqn_second}
\frac{\partial}{\partial t} N_m(t)
= -rN_m+\frac{2}{\hbar}\sum_n\mathcal{J}_{mn}\text{Im}[Z_{mn}]+\frac{2}{\hbar}\sum_nK_{mn}\text{Im}[R_{mn}]
\end{equation}

May and co-workers\cite{may2014kinetic} reduced the full master equation to kinetic rate equations by considering coherence between first excited states $W_{mn}=\langle B_m^{\dagger}B_n\rangle$, coherence between second excited state $Z_{mn}=\langle D_m^{\dagger}D_n\rangle$ and mixed coherence $R_{mn}=\langle D_m^{\dagger}B_n\rangle$ to decay rapidly, leading to 
\begin{align}
\label{eq:rate_description}
\frac{\partial}{\partial t} P_m(t)
&= -\kappa P_m + r N_m 
+\frac{2}{\kappa \hbar^2}\sum_n J_{mn}^2 [P_m N_n - P_n N_m] \nonumber \\
&\quad -\frac{2}{(\kappa +r)\hbar^2}\sum_n \mathcal{J}_{mn}^2 [P_m N_n - N_m P_n] \nonumber \\
&\quad -\frac{2}{(\kappa +r/2)\hbar^2}\sum_n K_{mn}^2 [P_m P_n + N_m N_n + P_n N_m - N_m] \nonumber \\
&\quad -\frac{2}{(\kappa +r/2)\hbar^2}\sum_n K_{nm}^2 [P_m P_n + N_m N_n + P_m N_n - N_n].
\end{align}
\begin{equation}
\begin{aligned}
\label{eq:rate_2nd_exc}
\frac{\partial}{\partial t} N_m(t)
&= - r N_m
+ \frac{2}{(\kappa +r)\hbar^2}\sum_n \mathcal{J}_{mn}^2
\bigl[ P_m N_n - N_m P_n \bigr] \\
&\quad
+ \frac{2}{(\kappa +r/2)\hbar^2}\sum_n K_{mn}^2
\bigl[ P_m P_n + N_m N_n + P_n N_m - N_m \bigr]
\end{aligned}
\end{equation}

The nonlinear rate equations describe multiexciton dynamics quite well in the weak-coupling limit where closed-system dynamics can be considered perturbative. Their validity arises not only from the vanishing time dependence of the coherences but also from retaining only the leading order contributions in $J$, $\mathcal{J}$, and $K$.

In this work, we extend May's approach to incorporate the effects of time-dependent off-diagonal elements originating from four-body correlations of transition operators.\cite{may2014kinetic} To close the resulting set of equations and keep the dynamics tractable, four-body correlators are factorized into pairwise transition operators or two-site correlation within a mean-field type approximation.\cite{wang2018laser} This procedure leads to the following simplified equation of motion for the off-diagonal elements\markchange{, with detailed derivation in the APPENDIX.}
\begin{widetext}
\begin{equation}
\begin{aligned}
\label{eq:single_exc_coh}
\frac{\partial}{\partial t} W_{mn}(t)
&= i\frac{(E_m-E_n)}{\hbar}W_{mn}-\kappa W_{mn}+\frac{iJ_{nm}}{\hbar}[(\Pi_m-P_m)P_n-(\Pi_n-P_n)P_m]\\
&\quad +\frac{i}{\hbar}\sum_{k\neq{n,m}} [J_{km}(\Pi_m-P_m)W_{kn}-J_{nk}(\Pi_n-P_n)W_{mk}]\\
&\quad +\frac{i}{\hbar}\sum_{k\neq{m}}\mathcal{J}_{mk}(Z_{mk}W_{mn}+R_{mn}R^*_{km})-\frac{i}{\hbar}\sum_{k\neq{n}}\mathcal{J}_{kn}(Z_{kn}W_{mn}+R^*_{nm}R_{kn})\\
&\quad +\frac{i}{\hbar}\sum_{k\neq{m}}K_{mk}(R_{mk}W_{mn}+R_{mn}W_{mk})-\frac{i}{\hbar}\sum_{k\neq{n}}K_{nk}(W_{mn}R^*_{nk}+R^*_{nm}W_{kn})\\
&\quad +\frac{i}{\hbar}\sum_{k\neq{m,n}}K_{km}(\Pi_m-P_m)R_{kn}-\frac{i}{\hbar}\sum_{k\neq{m,n}}K_{kn}(\Pi_n-P_n)R^*_{km}
\end{aligned}
\end{equation}
\begin{equation}
\begin{aligned}
\label{eq:double_exc_coh}
\frac{\partial}{\partial t} Z_{mn}(t)
&= i\frac{(\epsilon_m-\epsilon_n)-(E_m-E_n)}{\hbar}Z_{mn}-(\kappa +r)Z_{mn}-\frac{i}{\hbar}\sum_{k\neq{m}}J_{mk}(Z_{mn}W_{mk}+R_{mk}R^*_{nm})\\
& \quad +\frac{i}{\hbar}\sum_{k\neq{n}}J_{kn}(Z_{mn}W_{kn}+R_{mn}R^*_{nk}) +\frac{i}{\hbar}\mathcal{J}_{nm}[(P_m-N_m)N_n-(P_n-N_n)R_{mk})]\\
& \quad +\frac{i}{\hbar}\sum_{k\neq{m,n}}[\mathcal{J}_{km}(P_m-N_m)Z_{kn}-\mathcal{J}_{nk}(P_n-N_n)Z_{mk}] \\
&\quad +\frac{i}{\hbar}\sum_{k\neq{m,n}}[K_{mk}(P_m-N_m)R^*_{nk}-K_{nk}(P_n-N_n)R_{mk}] \\
&\quad -\frac{i}{\hbar}\sum_{k\neq{m}}K_{km}(R^*_{nm}Z_{mk}+Z_{mn}R^*_{km})+\frac{i}{\hbar}\sum_{k\neq{n}}K_{kn}(Z_{kn}R_{mn}+R_{kn}Z_{mn})
\end{aligned}
\end{equation}
\begin{equation}
\begin{aligned}
\label{eq:cross_coh}
\frac{\partial}{\partial t} R_{mn}(t)
&= i\,\frac{\left(\epsilon_m - E_m - E_n\right)}{\hbar}\,R_{mn}
 - \left(\kappa  + \frac{r}{2}\right) R_{mn}  \\
&\quad - \frac{i}{\hbar} \sum_{k \neq m} J_{mk}
\left( R_{mk} W_{mn} + R_{mn} W_{mk} \right) \\
&\quad - \frac{i}{\hbar} \sum_{k \neq m,n} J_{nk}\, R_{mk} \left( \Pi_n - P_n \right)
 + \frac{i}{\hbar} \sum_{k \neq m,n} \mathcal{J}_{km}\, R_{kn} \left( P_m - N_m \right) \\
&\quad - \frac{i}{\hbar} \sum_{k \neq n} \mathcal{J}_{kn}
\left( R_{mn} Z_{kn} + Z_{mn} R_{kn} \right) \\
&\quad + \frac{i K_{mn}}{\hbar}
\left[ (P_m - N_m) P_n - (\Pi_n - P_n) N_m \right] \\
&\quad + \frac{i}{\hbar}
\sum_{k\neq{m,n}}K_{mk}(P_m - N_m)W_{kn} - \frac{i}{\hbar}\sum_{k\neq{m,n}}K_{nk}(\Pi_n - P_n)Z_{mk} \\
&\quad -\frac{i}{\hbar}\sum_{k\neq{m}}K_{km}(R^*_{km}R_{mn}+Z_{mk}W_{mn})-\frac{i}{\hbar}\sum_{k\neq{n}}K_{nk}(R^*_{nk}R_{mn}+Z_{mn}W_{kn})
\end{aligned}
\end{equation}
\end{widetext}
These equations are the primary results of this paper and are employed in combination with the equation of motion of the population terms. {\markchange{To solve the equations of motion, Eq.~\eqref{eq:population_eqn} and Eq.~\eqref{eq:pop_eqn_second}, together with the coherence equations Eq.~\eqref{eq:single_exc_coh}, Eq.~\eqref{eq:double_exc_coh}, and Eq.~\eqref{eq:cross_coh}, we use the completeness relation [Eq.~\eqref{eq:completeness}] to reduce the number of independent variables.}}
In the following we shall examine a number of representative cases in various parameter regimes. 

\section{Initial coherent biexciton states}
To showcase the utility of this framework, we consider a one-dimensional molecular aggregate consisted of three-level systems in standard Frenkel exciton manners depicted in Fig.~\ref{fig:fig1}. We are interested in the spatiotemporal development of biexcitons that may result from realistic microscopy experiments.\cite{Libai2013ARMR,Zanni2016ACSPhotonics,Libai2017ACR,blach2025environment} Analyzing such trajectories on a lattice model can potentially correlate insights from time-resolved microscopy, typically carried out in the low fluence limit, with those obtained in more traditional EEA experiments where the focus is the transient non-exponential component of the fluorescence emission signal.

For the purpose of studying microscopy, we start with two normalized single-exciton wavepackets 
\begin{equation}
\sum_m |\phi^{(i)}_m|^2 = 1,
\qquad i=1,2,
\end{equation}
where the site index runs
\( m = 1,\dots,N_{mol} \). \markchange{The amplitudes $\phi_m^{(i)}$ are defined within the single-exciton manifold ($e$), where each wavepacket represents a single exciton delocalized over lattice sites. Therefore, they are indexed only by the site index $m$ and the wavepacket label $i=1,2$, without an additional electronic state index.

In our construction, the initial biexciton state is formed as an anti-symmetrized product of two single-exciton wavepackets, and thus the amplitudes $\phi_m^{(i)}$ correspond specifically to the $e$ state. The higher excited state $f$ is not directly populated in the initial condition, but rather becomes relevant dynamically through exciton–exciton fusion processes during time evolution.} The spatial overlap between the two single-exciton wavepackets is defined as
\begin{equation}
S = \langle \phi^{(1)} | \phi^{(2)} \rangle
= \sum_m \phi^{(1)*}_m \phi^{(2)}_m ,
\end{equation}
which characterizes the degree of spatial and momentum overlap between the two
excitons. Frenkel excitons are composite bosons subject to a hard-core constraint that forbids double occupancy of a site. This constraint can be enforced explicitly by using an anti-symmetrized biexciton wavefunction.
The anti-symmetrized and normalized biexciton wavefunction is given by
\begin{equation}
\Psi_{mn}
=
\frac{
\phi^{(1)}_m \phi^{(2)}_n
-
\phi^{(1)}_n \phi^{(2)}_m
}{
\sqrt{2\left(1-|S|^2\right)}
},
\end{equation}
where \( m \) and \( n \) label the lattice sites occupied by the two excitons.
The normalization factor ensures
\begin{equation}
\sum_{m,n} |\Psi_{mn}|^2 = 1.
\end{equation}

This biexciton wavefunction represents a pure biexciton state composed of two
indistinguishable excitons prepared in the wavepackets
\( \phi^{(1)} \) and \( \phi^{(2)} \).
For vanishing overlap \( S \approx 0 \), the state reduces to two well-separated
excitons, whereas finite overlap introduces exchange-induced quantum coherence
between the excitons. Such composite wavepackets can be thought as idealized excited state absorption contribution in a pump-probe microscopy measurement. 
\markchange{Although our three level aggregate's model includes a higher excited state $f$ ($S_n$ electronic state), this does not imply that two independent $e$ excitons ($S_1$) can occupy the same molecular site. Instead in our model, when two excitons interact, they undergo exciton fusion (annihilation), resulting in a transition of the form $S_1 + S_1 \rightarrow S_n + S_0$. Thus, double excitation on a single molecule corresponds to a different electronic configuration ($f$), rather than a product state of two $e$ excitations. Consequently, the local Hilbert space excludes double occupancy of the $e$ state, justifying the use of hard-core 
bosons. This treatment is consistent with the standard description of Frenkel excitons, where excitations behave as bosons on different sites but are subject to Pauli exclusion locally. The anti-symmetrized structure therefore enforces the exclusion of multiple occupancy of the same excitonic state, rather than implying true fermionic statistics.\cite{spano1991fermion} 

\markchange{This should be contrasted with recent soft-core approaches,\cite{bubilaitis2024signatures,bubilaitis2025compact} where multiple occupancy is allowed intrinsically through modified commutation relations or finite on-site interaction energies. In our framework, the hard-core nature of $e$ excitons is strictly preserved, and higher excitation effects arise only through explicit coupling to the separate $f$ manifold. Therefore, our model can be viewed as a complementary physical limit to these soft-core descriptions, where multi-exciton effects are generated dynamically via fusion processes rather than through relaxed occupancy constraints.}}

To characterize observables associated with single-exciton operators, we
construct the one-particle reduced density matrix from two-particle description by tracing over one exciton,
\begin{equation}
\rho_{mn}
=
\sum_q \Psi_{mq}\Psi^*_{nq}.
\end{equation}
Substituting the anti-symmetrized biexciton wavefunction yields
\begin{equation}
\rho_{mn}
=
\frac{
\phi^{(1)}_m \phi^{(1)*}_n
+
\phi^{(2)}_m \phi^{(2)*}_n
-
S\,\phi^{(1)}_m \phi^{(2)*}_n
-
S^*\,\phi^{(2)}_m \phi^{(1)*}_n
}{
1-|S|^2
}.
\end{equation}

The diagonal elements of the reduced density matrix,
\begin{equation}
P_m = \rho_{mm},
\end{equation}
represent the average exciton population at site \( m \), while the off-diagonal
elements describe single-exciton quantum coherence.
The trace of the reduced density matrix satisfies
\begin{equation}
\mathrm{Tr}\,\rho = \sum_m \rho_{mm} = 2,
\end{equation}
reflecting the presence of two excitons in the biexciton state. As we are interested in the dependence of initial spatial profiles, in the context of optical microscopy, we assume each of the constituent single exciton as a standing wave under a Gaussian envelope, whose wavefunctions can be written as
\begin{equation}
\phi^{(i)}_m
=
\frac{\sqrt{2}}{\sqrt{w\sqrt{\pi}\left(1+e^{-q_i^2 w^2}\cos\left(2q_im_{0i}\right)\right)}}
\cos\left(q_i m\right)\,
\exp\!\left[-\frac{(m-m_{0i})^2}{2w^2}\right],
\end{equation}
where $q_i=\frac{\pi k_i}{N_{mol}+1}$ and \( m_{0i} \) denote the spatial centers of the wavepackets,
\( w \) is their spatial width, and \markchange{ $k_i$ is a positive integer representing central
momenta.}
Similarly, considering each of the initial wavepacket as traveling gaussian wave, the single exciton wavefunction can be expressed as
\begin{equation}
\phi^{(i)}_m
=
\frac{1}{\sqrt{w\sqrt{\pi}}}
\exp\!\left[
-\frac{(m-m_{0i})^2}{2w^2}
+ i q_i m
\right]
\end{equation}

\markchange{While it is true that in ideal J- and H-aggregates only the $k=1$ exciton state dominates the linear absorption. In real-space excitation, such as in time-resolved optical microscopy or pump–probe experiments, the excitation is spatially localized, which naturally generates a superposition of exciton eigenstates rather than a single momentum eigenstate. This leads to the formation of exciton wavepackets with finite spatial extent and momentum distribution.

Furthermore, the biexciton states considered in our work can be accessed via excited-state absorption (ESA) from the initially prepared bright single-exciton state. The optical selection rules govern these transitions,\cite{spano1991fermion} allowing access only to specific biexciton states [e.g., $(2,N)$ in H-aggregates] that satisfy wavevector matching and parity conditions. Thus, even though only one single-exciton state is directly bright, the combination of spatially localized excitation and ESA pathways enables the preparation of coherent biexciton wavepackets. In contrast, the incoherent initial conditions used in our calculations are idealized and serve as a reference point to the available nonlinear kinetic theories.\cite{gaivzauskas1995annihilation,malyshev1999exciton,ryzhov2001low}}
\section{Result and discussion}
We are interested in the features of biexciton dynamics in time-resolved fluorescence emission and correlate them with those in its spatio-temporal development. The former can be defined in terms of decay from the first and second excited states as follows
\begin{equation}
I\left(t\right)=\sum_n[\kappa P_n(t)+rN_n(t)]
\end{equation}
Total exciton number is defined as
\begin{equation}
N_{tot}\left(t\right)=\sum_n[P_n(t)+2N_n(t)]
\end{equation}
While a common coarse-grained metric of the latter is the mean squared displacement (MSD) that can be obtained using the following equation
\begin{equation}
\mathrm{MSD}(t)
=
\frac{\sum_{m=1}^{N_{\mathrm{mol}}}
\bigl(m-\bar{m}(t)\bigr)^2
\left[P_m(t)+2N_m(t)\right]}
{\sum_{m=1}^{N_{\mathrm{mol}}}
\left[P_m(t)+2N_m(t)\right]} ,
\end{equation}

\begin{equation}
\bar{m}(t)
=
\frac{\sum_{m=1}^{N_{\mathrm{mol}}}
m\left[P_m(t)+2N_m(t)\right]}
{\sum_{m=1}^{N_{\mathrm{mol}}}
\left[P_m(t)+2N_m(t)\right]} .
\end{equation}

Time derivative of MSD provides the time dependent diffusivity in one dimension as follows
\begin{equation}
D_t=\frac{1}{2}\frac{d}{dt}[\text{MSD}(t)]
\end{equation}
which in the diffusive regime becomes the diffusion constant. \markchange{The time-resolved emission and transport properties are computed for incoherent and coherent initial biexciton states in a one-dimensional open lattice chain with 21 three-level sites. The initial biexcitons (corresponding to the maxima of the wavepackets for coherent initial conditions) are placed symmetrically around the middle site of the chain. The equations of motion are solved using the \texttt{ode15s} solver in MATLAB, which employs variable-step implicit methods suitable for stiff systems arising from the nonlinearity of exciton–exciton annihilation and the presence of widely separated timescales. The time step used for propagation is $10^{-2}$ in units of $J^{-1}$. Identical dynamics obtained with a smaller time step of $10^{-3}$ (in units of $J^{-1}$) confirm the convergence of the results.}

\subsection{Incoherent initial biexciton states}
We begin our analysis by using an incoherent rate-equation framework Eq.~\eqref{eq:rate_description} to describe experimentally accessible quantities including time-resolved emission and exciton density diffusion. Furthermore, we assume  the initial state of the dynamics to be a fully localized, double Kronecker delta $\phi^{(1)}_m=\delta_{n-1,m}$ and $\phi^{(2)}_m=\delta_{n+1,m}$. Although such description is inadequate for describing coherent effect and a completely localized initial double excitation is generally not experimentally realizable in molecular aggregate context, it offers a simplified approach to elucidate the competition among different processes in multilevel molecular aggregate systems.
\begin{figure*}
 \centering
 \includegraphics[height=10cm]{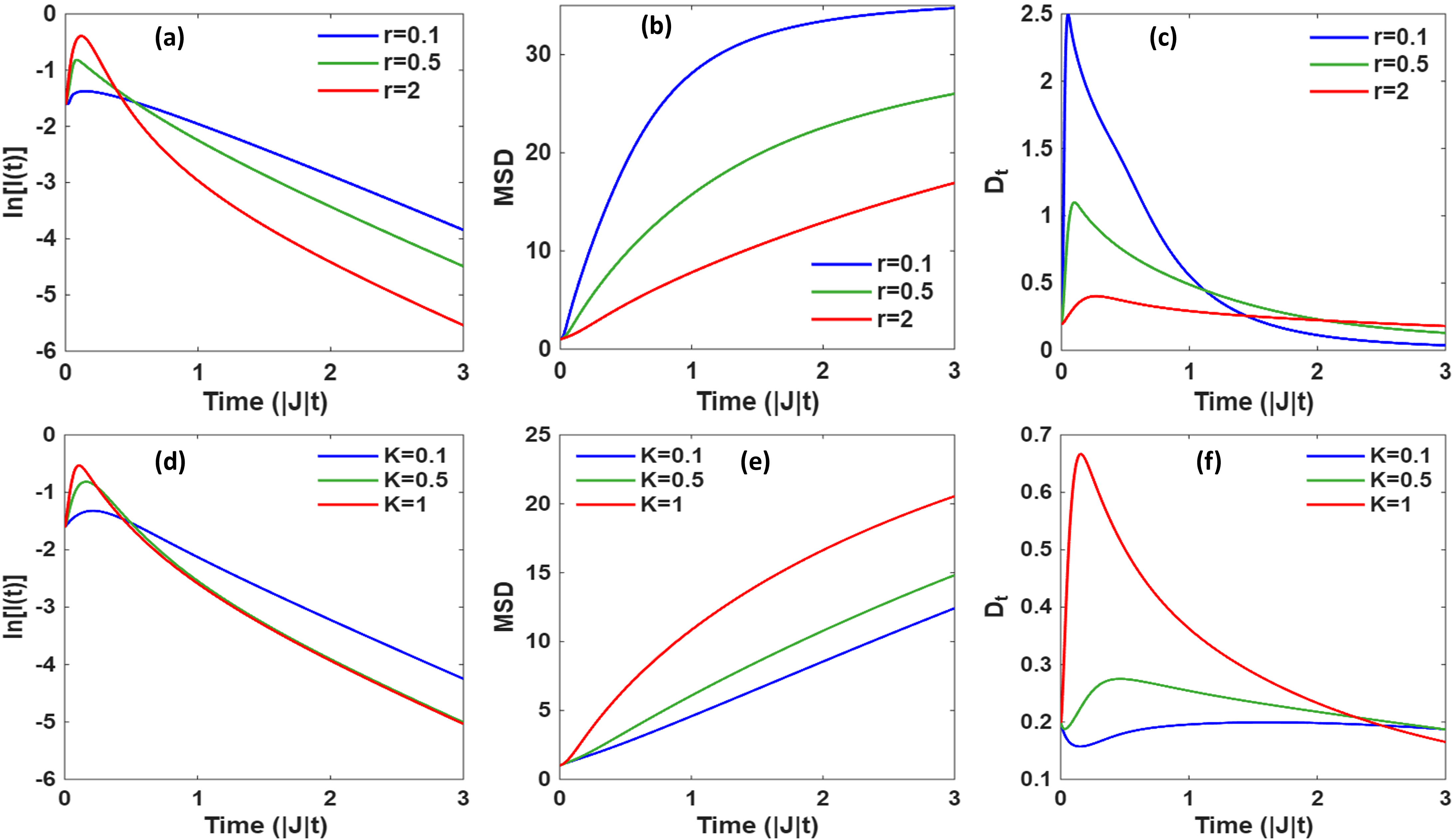}
\caption{For a linear chain of three-level systems with nearest neighbor interaction with $J=\mathcal{J}=-0.1$ and decay from first excited state $\kappa =0.1$ semi-log plot of time dependent emission, mean squared displacement and time dependent diffusivity are plotted, with varying decay rate $r$ from second excited state in panels (a) to (c) at fixed $K=1$ and with varying fusion $K$ at fixed decay rate $r=1$ in panels (d) to (f). Initially biexcitons are placed symmetrically with respect to the center of the 21-site molecular aggregates. The dynamics is propagated with population-only kinetic equations. {\markchange{The time axis is scaled by the absolute value of the nearest-neighbor intermolecular Coulomb coupling ($J$) in the single-exciton manifold.}} }
 \label{fig:fig2}
\end{figure*}
In Fig.~\ref{fig:fig2}, the time-dependent emission intensity, MSD, and time-dependent diffusivity are shown for different fusion and recombination rates initialized from the double Kronecker delta state. 
Although the dynamics is governed by incoherent rate equations, the emission exhibits a pronounced non-exponential decay. This non-exponential behavior arises from the nonlinear coupling between exciton and biexciton populations, together with transport-induced time-dependent decay rates, rather than from quantum coherence effects. 

Further scrutinizing the results obtained from the incoherent description, we note that the interconversion between the populations $P_m$ and $N_m$ is controlled by the fusion rate $K$, while the non-radiative decay of the fusion product is described by the recombination rate $r$. The resulting relaxation is inherently non-exponential, reflecting the coexistence of multiple inter-converting populations and the spatial profile-dependent nature of EEA. At early times, the high exciton density of the first-excited state enhances fusion, rapidly transferring the population from the two-exciton manifold $P_m$ into the fusion channel $N_m$, leading to a rapid initial decay of the total excitation signal. As the system evolves, spatial spreading reduces local excitation density, suppressing further fusion, and slowing the effective decay rate, which manifests as pronounced curvature in $\ln[I(t)]$.
This interplay is strongly modulated by the lifetime of the fusion product. For small recombination rates $r$, the $N_m$ population persists and temporarily stores excitation density, allowing continued spatial redistribution and resulting in a larger MSD and enhanced time dependent diffusivity $D_t$. In contrast, larger values of $r$ rapidly remove fusion products, limiting the time available for spatial spreading and reducing both the MSD and $D_t$. Similarly, increasing the fusion strength $K$ accelerates the depletion of the two-exciton population $P_m$, suppressing long-range transport despite the absence of quantum coherence.\cite{tutunnikov2023coherent} For small $K$, fusion is inefficient at initial times, allowing the two-exciton population to persist and spread spatially, which leads to rapid MSD growth and initially increasing $D_t$. Increasing $K$ enhances early-time annihilation, producing an early maximum in $D_t$ followed by monotonic decay. 

\subsection{Coherent initial biexciton states}

\begin{figure*}
 \centering
 \includegraphics[height=10cm]{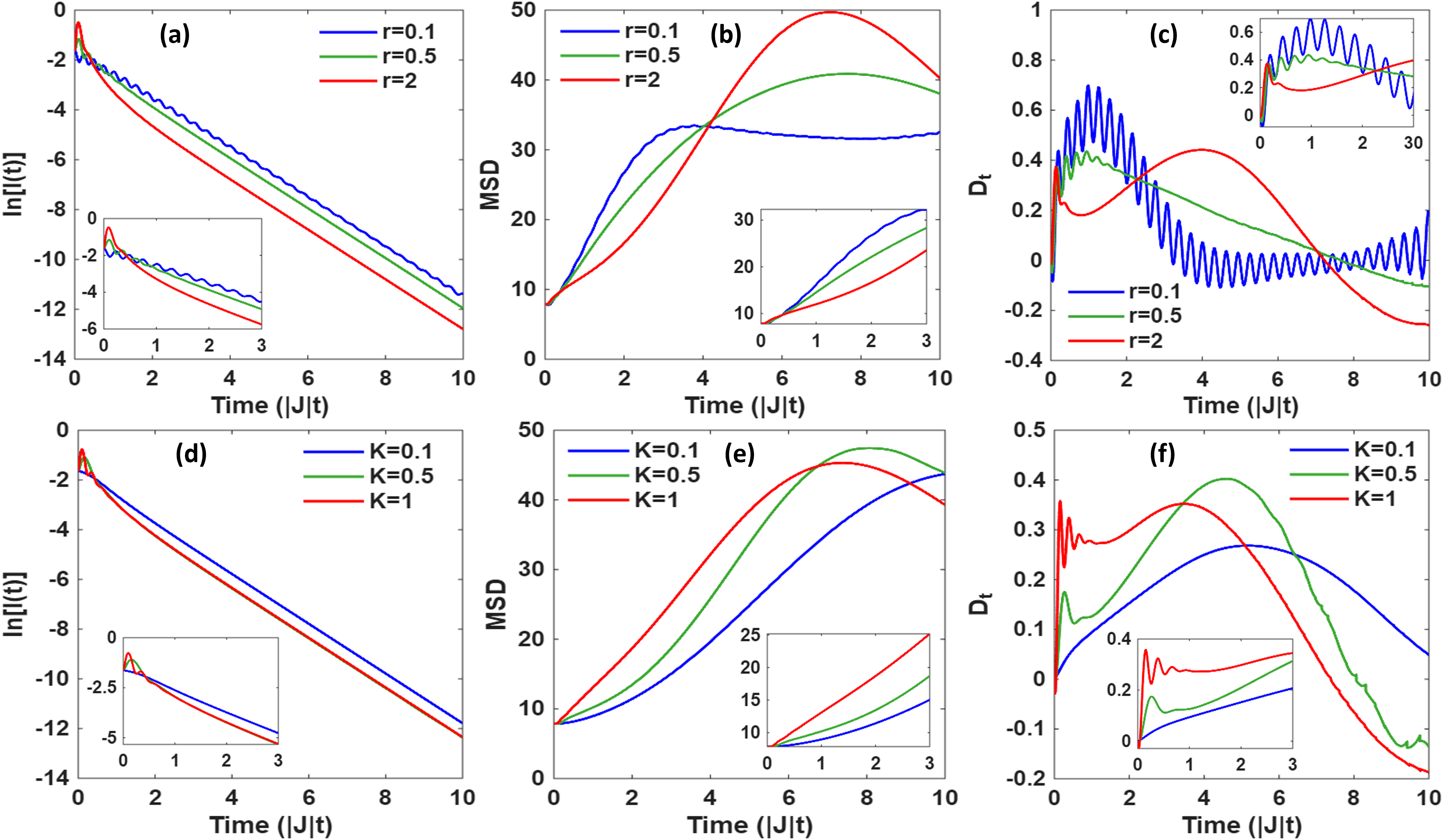}
\caption{
Same system specifications as in Fig.~\ref{fig:fig2} but for initial biexciton state corresponding to two exciton gaussian wavepackets with $k_1=1$ and $k_2=2$ with the width $w=3$ with the centre of wavepackets at $m_{01}=10$ and $m_{01}=12$, described in main text. The dynamics is propagated with fully coherent description. The inset highlights the short-time MSD and $D_t$. {\markchange{The time axis is expressed in units of inverse of the absolute value of the nearest-neighbor Coulomb coupling ($J$) in the single-exciton manifold.}}
}
 \label{fig:fig3}
\end{figure*}
To develop experimentally relevant biexciton initial states and examine their impact on the interplay between quantum coherence and EEA dynamics, we first investigate biexcitons initially prepared as standing-wavepackets (Fig.~\ref{fig:fig3}). The corresponding wave vectors are chosen such that these states represent optically bright transitions from the band edge of the single-exciton manifold to the biexciton manifold in a J-aggregate.\cite{spano1991fermion} By varying the width of the wavepacket $w$, the initial state can be tuned from a localized, effectively incoherent regime in small $w$ to a fully delocalized regime in large $w$. We consider that the population and coherence are initially restricted to the first excited states as a biexciton. The inclusion of coherence in the equation of motion [Eq.~\eqref{eq:single_exc_coh} to Eq.~\eqref{eq:cross_coh})] together with a coherently prepared biexciton wavepacket produces pronounced transient features (Fig.~\ref{fig:fig3}) that are absent in the incoherent kinetic description. The initial biexciton state is prepared as an anti-symmetrized product of two spatially extended coherent single-exciton wavepackets both centered at the middle of the chain, which gives rise to nonzero off-diagonal site coherences in the reduced density matrix. These coherences encode phase relations between different sites, leading to transient coherent spreading that suppresses fusion and recombination, giving rise to shoulders and curvature in the emission decay as well as the enhanced early-time MSD growth. (Insets of Fig.~\ref{fig:fig3}, where the time frame is the same as in Fig.~\ref{fig:fig2}.) While incoherent rate equations predict purely diffusive spreading, the inclusion of quantum coherences and phase-correlated initial wavepackets reveals a pronounced ballistic transport regime at early times (MSD$\propto t^2$ and $D_t\propto t$). 
Non-trivial temporal and spatial profiles last until after coherences decay, and the dynamics cross over to a purely kinetic regime similar to that described by rate equations. Since the biexciton population and its associated coherences decay slowly for small $r$, population–coherence feedback loops persist over long times, leading to oscillatory transport and emission dynamics. The coherent-to-incoherent transition and the duration of diffusive spread of time-dependent exciton density are sensitive to the exciton fusion rate ($K$) and the decay to the first excited state ($r$). This demonstrates that, even within a Markovian framework, coherent state preparation fundamentally reshapes the interplay between transport, manifold-crossing, and annihilation, yielding relaxation dynamics that cannot be inferred from population kinetics alone. 

Furthermore, as we follow the trajectories in Fig.~\ref{fig:fig3} to longer times, the wavepackets reach the open boundaries of the chain and bounce back, evidenced by the shrinking of MSD and negative $D_t$ values. This can be considered as a crude approximation to disorder where the excitons are forced to localize within a infinite square well. On the other hand, no such information can be deduced from the time-resolved fluorescence signal. This again demonstrates the limitation of the latter as a means to gauge transport. \markchange{The condition $\epsilon_m = 2E_m$ used in our calculations corresponds to a resonant situation for exciton–exciton fusion, where excitons can efficiently access the higher excited state. In realistic systems, this condition is generally not satisfied exactly, leading to a finite detuning. In the presence of such detuning, the fusion process becomes off-resonant, resulting in a reduced transition rate to the higher excited state. Consequently, exciton–exciton annihilation is suppressed, leading to a slower decay of the biexciton population. As a result, signatures of the initial coherent state can persist for longer times, and coherent transport effects may be sustained over extended timescales.}

\subsection{J- vs H-aggregates}
\begin{figure*}
 \centering
 \includegraphics[height=10cm]{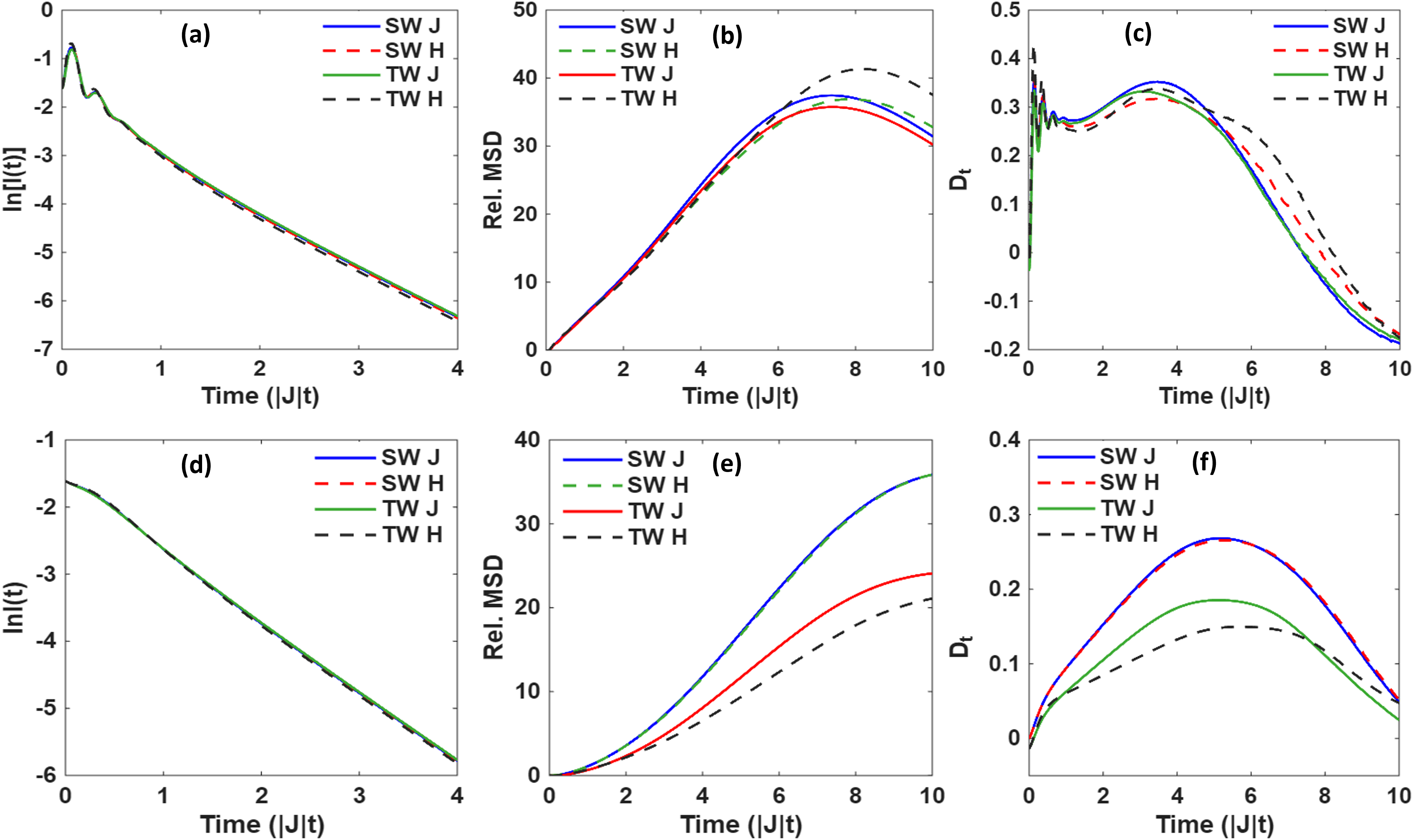}
\caption{
Same setup as in Fig.~\ref{fig:fig3} with $r=1$. Panels (a) to (c) are for fixed $K=1$ and in panels (d) to (f)  $K=0.1$. Here the initial biexciton states are standing wave (SW) packets with $k_1=1$ and $k_2=2$ and traveling wave (TW) packets with $k_1=2$ and $k_2=N_{mol}$ for J and H aggregate, respectively.
 {\markchange{The time axis is given in terms of inverse of the absolute value of the nearest-neighbor Coulomb coupling ($J$) in the single-exciton manifold.}}}
 \label{fig:fig4}
\end{figure*}

An interesting question that arises is whether one can distinguish the spatio-temporal dynamics of biexcitons given the exciton band structure. As is well known, for an incoherent initial condition in the single-exciton manifold, J- and H-aggregates cannot be distinguished through population dynamics alone. To separate their effects, one must instead examine coherences or related optical properties such as absorption and emission spectra. Similarly, for incoherent biexciton initial conditions, J- and H-aggregates cannot be distinguished using time-resolved emission or MSD, as both probes only the population dynamics. To compare J- and H-aggregates, instead of MSD, we consider relative MSD, defined as $\text{MSD}(t)-\text{MSD}(0)$, since the initial spread of exciton density differs for J- and H-aggregates due to their different momentum components.
In Fig.~\ref{fig:fig4} the distinction between J- and H-aggregates is determined not only by the \textit{sign} of the excitonic coupling but also by the momentum composition of the optically bright biexciton state. For J-aggregates, the initial biexciton is constructed from the lowest two standing-wave exciton modes $k_1=1$ and $k_2=2$, whereas for H-aggregates it involves high-momentum modes near the top of the exciton band at $k_1=2$ and $k_2=N_{mol}$.\cite{spano1991fermion}  

\markchange{The choice of the biexciton bright state follows from the optical selection rules governing one- to two-exciton transitions. For H-aggregates the bright states from the band edge $k=N_{mol}$ state is dominated by the product state $(k_1,k_2) = (2, N_{mol})$, among other weaker contributing product states. The transition dipole moment depends on the overlap between a single-exciton wavefunction and an anti-symmetrized biexciton wavefunction.\cite{spano1991fermion} This overlap is non-zero only when specific conditions are satisfied, including a resonance condition $k = \pm k_1 \pm k_2$ (mod $N_{mol}+1$) and a parity matching condition, requiring that $k$ and $k_1 + k_2$ have the same parity (both even or both odd). For an H-aggregate, the initial bright state of single exciton corresponds to $k = 1$, and the band edge state is $k = N_{mol}$. The transition from the latter to $(k_1,k_2) = (1, N_{mol})$ is forbidden due to parity mismatch, while $(2, N_{mol})$ satisfies both the resonance and parity conditions, making it the lowest-order optically allowed biexciton state. Higher-order states such as $(4, N_{mol})$, $(6, N_{mol})$, etc., are also symmetry-allowed but exhibit progressively weaker intensities. This reduction arises from the introduction of more nodes in the wavefunctions, which leads to partial cancellation in the overlap integral and hence a smaller transition dipole moment.}

 Although these states are all optically bright and exhibit nearly identical emission dynamics, their transport properties differ markedly. The MSD and the time-dependent diffusivity exhibit a pronounced sensitivity to the momentum composition and phase structure of the initial biexciton state. This sensitivity results in enhanced spatial spreading in J-aggregates and suppressed transport in H-aggregates, particularly under traveling-wave initialization. This behavior can be understood from the population current appearing in Eq.~\eqref{eq:population_eqn}, which is proportional to the imaginary part of the inter-site coherences. For a homogeneous molecular aggregate with nearest-neighbor coupling $J$, the exciton dispersion ($2J\cos k$) results a group velocity ($2J\sin k$). Since the population current is directly proportional to imaginary part of the coherences, it is therefore linked to both the magnitude and \textit{sign} of $2J\sin k$, and hence to the phase gradients encoded in the coherences.

For standing-wave initial conditions, the biexciton state contains equal contributions from $+k$ and $-k$ components. These opposite momenta generate currents of equal magnitude but opposite sign, leading to cancellation of the net current at $t=0$. Consequently, the initial imaginary part of the coherences vanishes, and there is no net population current at the time of initial state preparation. Since there is no initial imaginary component of the coherences, the dynamically generated imaginary part that develops during the evolution is not sufficient to distinguish between J- and H-aggregates. Consequently, the transport dynamics under standing-wave are significantly less sensitive to the \textit{sign} of the coupling and to the momentum ($k$) component compared to the traveling-wave case, resulting in similar transport behavior for J- and H-aggregates.

In contrast, traveling-wave initial conditions carry a definite momentum and thus possess a finite imaginary component of the coherences and a nonzero population current. The subsequent dynamics therefore become sensitive to both the sign of the coupling and the value of the momentum of the initially prepared optically bright states. In J-aggregates, the biexciton is predominantly formed from low-order exciton modes located near the bottom of the exciton band, where the dispersion is steep. The large slope of the dispersion corresponds to a large group velocity, so phase gradients efficiently generate population currents. Traveling-wave preparation thus produces a substantial imaginary component of the intersite coherences, resulting in enhanced spatial spreading and a larger MSD. In contrast, the H-aggregate biexciton contains significant contributions from modes close to the upper band edge (large $k$ states), where the dispersion becomes locally flat and the group velocity is strongly reduced. Because the population current scales with $2J\sin k$, these band edge components generate only weak current-carrying coherences. Even under traveling-wave initialization, the phase evolution produces only a small imaginary part of the coherences and a correspondingly weak net population current. As a result, spatial transport is suppressed and the MSD remains small. As a result, transport is suppressed in H-aggregates. These differences highlight the role of coherence and band-dispersion effects in biexciton transport, which are completely absent in incoherent rate-equation descriptions.



\section{Conclusions}
In this work, we have developed an extended theoretical framework for describing biexciton dynamics in three-level molecular aggregates that goes beyond phenomenological rate-equation models by explicitly incorporating populations, single and biexciton manifold coherences, and cross-manifold coherences within a single dynamical description. By extending May’s kinetic approach\cite{may2014kinetic} into the regime of intermediate and strong inter-site coupling, we have shown that coherences between the single- and double-excitation manifolds cannot, in general, be neglected and play a decisive role in determining both the temporal relaxation and spatial transport of multiexciton states. The resulting equations of motion provide a transparent yet flexible platform for analyzing exciton–exciton annihilation in systems where delocalization, coherence, and dissipation coexist.

A key finding of this study is that the nature of the initial biexciton state and its spatial extent, momentum composition, and phase coherence could reshape the subsequent dynamics. In the case of incoherent initial conditions, the dynamics is governed by nonlinear population kinetics arising from the coupling between single and double excitation manifolds. This results in a strong non-exponential emission decay reflecting the interplay between fusion and recombination. Even in this incoherent limit, the spatial spreading of the exciton density cannot be described by a simple bimolecular rate law due to the presence of exciton annihilation which couples the single and double excitation.
When coherently prepared biexciton states are considered, qualitatively different and novel dynamical features emerge. Coherent initial wavepackets generate pronounced transient dynamical features in both emission and transport, including early-time ballistic spreading and a subsequent slowdown of the dynamics, arising from the suppression of effective annihilation and decay compared with incoherent predictions. These effects originate from nonvanishing coherences in the purely single-exciton and biexciton manifolds, as well as cross-coherences between the two manifolds of the reduced density matrix, which encode phase correlations between spatially separated sites and give rise to coherence-induced transport flux. As these coherences decay, the system crosses over to a kinetic regime, demonstrating how coherent and incoherent dynamics are continuously connected within the present framework rather than representing distinct limiting cases.

Another important conclusion of this work is the identification of clear signatures that distinguish J- and H-aggregates at the biexciton level. While population-based observables such as time-resolved emission remain largely insensitive to the \textit{sign} of the excitonic coupling and momentum composition to the bright biexciton states under coherent initial state preparation, transport observables remain strongly sensitive to the momentum content and phase structure of the biexciton wavepacket, especially for initial states prepared as traveling wavepackets. For coherently prepared states, J-aggregates exhibit enhanced spreading driven by initial net population current arising from low-momentum exciton modes with steep dispersion and finite group velocities, whereas H-aggregates show suppressed transport because band-edge modes with locally flat dispersion generate weak current-carrying coherences. This demonstrates that biexciton transport provides a powerful probe of aggregate type and band structure that is inaccessible through time-resolved emission measurements alone.

Beyond the specific results presented here, the theoretical formulation developed in this work opens several avenues for future investigation. The present approach can be systematically extended to include higher multiexciton manifolds, enabling fluence dependence in nonlinear spectroscopic experiments. Incorporating non-Markovian environmental effects could be another important extension, since memory induced system-bath correlations may sustain long-lived coherences and alter annhilation pathways in ways that cannot be captured within a purely Markovian description. 

In summary, we have established a coherent description of biexciton dynamics that unifies transport, annihilation, and optical relaxation within a single theoretical framework. Our results highlight quantum coherence not only as a temporal effect, but also shaped by the spatial structure of the wavepacket that can be harnessed to alter multiexciton processes in complex excitonic systems. By demonstrating how coherence, band structure, and initial-state preparation jointly control biexciton transport and annihilation, this work offers a microscopic basis for designing materials and excitation protocols that reduce annihilation losses while preserving efficient energy transport or probing microscopic details that are otherwise challenging to interrogate.
\appendix
\section{Derivation of Equation of motion for population and coherences}
To extract population and coherences from the Lindblad equation [Eq.\eqref{eq:lindblad}] along with dissipator equations [Eq.\eqref{eq:dissipator_1} and Eq.\eqref{eq:dissipator_2}], we define the expectation value of an arbitrary operator $\hat{O}$ as
\begin{equation}
\operatorname{tr}\{\hat{\rho}(t)\,\hat{O}\} = \left\langle\hat{O}\right\rangle
\end{equation}
Here, $\hat{O}$ is formed by the transition operators $B_m^\dagger$($B_m$) and $D_m^\dagger$ ($D_m$). Transition operators are defined as
\begin{equation}
\begin{aligned}
B_m^{\dagger} &= \lvert \phi_{me} \rangle \langle \phi_{mg} \rvert, \\
B_m     &= \lvert \phi_{mg} \rangle \langle \phi_{me} \rvert.
\end{aligned}
\end{equation}

\begin{equation}
\begin{aligned}
D_m^{\dagger} &= \lvert \phi_{mf} \rangle \langle \phi_{me} \rvert, \\
D_m     &= \lvert \phi_{me} \rangle \langle \phi_{mf} \rvert.
\end{aligned}
\end{equation}

In the next step, we use the following commutation relation that simplifies the equation of motion as
\begin{align}
[B_m, B_n] &= 0, \quad [B_m^\dagger, B_n^\dagger] = 0, \quad m \neq n \\
[D_m, D_n] &= 0, \quad [D_m^\dagger, D_n^\dagger] = 0, \quad m \neq n \\
[B_m, B_m^\dagger] &= |\phi_{mg}\rangle\langle \phi_{mg}| - |\phi_{me}\rangle\langle \phi_{me}| \\
[D_m, D_m^\dagger] &= |\phi_{me}\rangle\langle \phi_{me}| - |\phi_{mf}\rangle\langle \phi_{mf}| \\
[B_m, D_m] &= B_m D_m =|\phi_{mg}\rangle\langle \phi_{mf}| \\
[D_m^\dagger, B_m^\dagger] &= D_m^\dagger B_m^\dagger = |\phi_{mf}\rangle\langle \phi_{mg}| \\
[B_m, D_m^\dagger] &= 0, \quad [B_m^\dagger, D_m] = 0
\end{align}

Here, we have used the orthonormality of the states. The above equations also indicate that transition operators on different sites commute with each other. In addition, we assume that direct transitions between the ground and excited states are not allowed, which implies that the following commutator also vanishes.
\begin{align}
[B_m, D_m] &= 0 \\
[D_m^\dagger, B_m^\dagger] &= 0
\end{align}

Now, the population equation of first excited states can be written as 
\begin{equation}
\begin{aligned}
\label{eq:pop_first_app}
\frac{\partial}{\partial t} \langle B_m^\dagger B_m \rangle
&= - \kappa  \langle B_m^\dagger B_m \rangle + r \langle D_m^\dagger D_m \rangle \\
&\quad - \frac{i}{\hbar} \sum_n \left( J_{mn} \langle B_m^\dagger B_n \rangle - J_{nm} \langle B_n^\dagger B_m \rangle \right) \\
&\quad - \mathcal{J}_{mn} \langle D_m^\dagger D_n \rangle + \mathcal{J}_{nm} \langle D_n^\dagger D_m \rangle \\
&\quad + K_{mn}^* \langle B_n^\dagger D_m \rangle - K_{mn} \langle D_m^\dagger B_n \rangle \\
&\quad + K_{nm}^* \langle B_m^\dagger D_n \rangle - K_{nm} \langle D_n^\dagger B_m \rangle
\end{aligned}
\end{equation}
Combining all the complex conjugate terms in Eq. \eqref{eq:pop_first_app} lead to the Population equation Eq. \eqref{eq:population_eqn}. Similarly the population equation for the second excited states can be written as
\begin{equation}
\begin{aligned}
\label{eq:pop_second_app}
\frac{\partial}{\partial t} \langle D_m^\dagger D_m \rangle
&= - r \langle D_m^\dagger D_m \rangle \\
&\quad +\frac{i}{\hbar} \sum_n (\mathcal{J}_{mn} \langle D_m^\dagger D_n \rangle - \mathcal{J}_{nm} \langle D_n^\dagger D_m \rangle) \\
&\quad  K_{mn}^* \langle B_n^\dagger D_m \rangle - K_{mn} \langle D_m^\dagger B_n \rangle
\end{aligned}
\end{equation}
Simplification by combining complex conjugate terms in Eq. \eqref{eq:pop_second_app} gives Eq. \eqref{eq:pop_eqn_second}.

Using the Lindblad equation [Eq.\eqref{eq:lindblad}] and the commutation relations equation for coherences between first excited, second excited levels and cross coherence can be expressed in terms of the expectation value product of multiple transition operator as
\begin{equation}
\begin{aligned}
\frac{\partial}{\partial t} \langle B_m^\dagger B_n \rangle
&= i\frac{(E_m-E_n)}{\hbar}\langle B_m^\dagger B_n \rangle-\kappa \langle B_m^\dagger B_n \rangle \\
&\quad + \frac{i}{\hbar} \sum_k \Big(
J_{km} \langle (B_m B_m^\dagger - B_m^\dagger B_m) B_k^\dagger B_n \rangle \\
&\qquad - J_{nk} \langle B_m^\dagger B_k (B_n B_n^\dagger - B_n^\dagger B_n) \rangle \\
&\qquad + \mathcal{J}_{mk} \langle D_m^\dagger B_m^\dagger D_k B_n \rangle 
- \mathcal{J}_{kn} \langle B_m^\dagger D_k^\dagger B_n D_n \rangle \\
&\qquad + K_{mk} \langle D_m^\dagger B_m^\dagger B_k B_n \rangle 
- K^*_{nk} \langle B_m^\dagger B_k^\dagger B_n D_n \rangle \\
&\qquad + K_{km} \langle (B_m B_m^\dagger - B_m^\dagger B_m) D_k^\dagger B_n \rangle \\
&\qquad - K^*_{kn} \langle B_m^\dagger D_k (B_n B_n^\dagger - B_n^\dagger B_n) \rangle
\Big)
\end{aligned}
\end{equation}
\begin{equation}
\begin{aligned}
\frac{\partial}{\partial t} \langle D_m^\dagger D_n \rangle
&= i\frac{(\epsilon_m-\epsilon_n)-(E_m-E_n)}{\hbar}\langle D_m^\dagger D_n \rangle-(\kappa +r) \langle D_m^\dagger D_n \rangle \\
&\quad - \frac{i}{\hbar} \sum_k \Big(
J_{mk} \langle D_m^\dagger B_m^\dagger B_k D_n \rangle 
- J_{kn} \langle D_m^\dagger B_k^\dagger B_n D_n \rangle \\
&\qquad - \mathcal{J}_{km} \langle (D_m D_m^\dagger - D_m^\dagger D_m) D_k^\dagger D_n \rangle \\
&\qquad + \mathcal{J}_{nk} \langle D_m^\dagger D_k (D_n D_n^\dagger - D_n^\dagger D_n) \rangle \\
&\qquad - K^*_{mk} \langle (D_m D_m^\dagger - D_m^\dagger D_m) B_k^\dagger D_n \rangle \\
&\qquad + K^*_{km} \langle D_k D_m^\dagger B_m^\dagger D_n \rangle \\
&\qquad + K_{nk} \langle D_m^\dagger B_k (D_n D_n^\dagger - D_n^\dagger D_n) \rangle \\
&\qquad - K_{kn} \langle D_m^\dagger B_n D_n D_k^\dagger \rangle
\Big)
\end{aligned}
\end{equation}
\begin{equation}
\begin{aligned}
\frac{\partial}{\partial t} \langle D_m^\dagger B_n \rangle
&= i\,\frac{\left(\epsilon_m - E_m - E_n\right)}{\hbar}\,\langle D_m^\dagger B_n \rangle
 - \left(\kappa  + \frac{r}{2}\right) \langle D_m^\dagger B_n \rangle \\
&\quad - \frac{i}{\hbar} \sum_k \Big(
J_{mk} \langle D_m^\dagger B_m^\dagger B_k B_n \rangle \\
&\qquad + J_{nk} \langle D_m^\dagger (B_n B_n^\dagger - B_n^\dagger B_n) B_k \rangle \\
&\qquad - \mathcal{J}_{km} \langle D_k^\dagger (D_m D_m^\dagger - D_m^\dagger D_m) B_n \rangle \\
&\qquad + \mathcal{J}_{kn} \langle D_m^\dagger D_k^\dagger B_n D_n \rangle \\
&\qquad - K^*_{mk} \langle (D_m D_m^\dagger - D_m^\dagger D_m) B_k^\dagger B_n \rangle \\
&\qquad + K^*_{kn} \langle D_m^\dagger D_k (B_n B_n^\dagger - B_n^\dagger B_n) \rangle \\
&\qquad + K^*_{km} \langle D_k D_m^\dagger B_m^\dagger B_n \rangle \\
&\qquad + K^*_{nk} \langle D_m^\dagger B_k^\dagger B_n D_n \rangle
\Big)
\end{aligned}
\end{equation}

The expectation value of product of multiple transition operator can be written as product of population and coherences using the following mean field type approximation.

\begin{equation}
\label{mean_field}
    \langle ABCD \rangle=\langle AB\rangle \langle CD\rangle + \langle AC\rangle \langle BD\rangle + \langle AD\rangle \langle BC\rangle
\end{equation}

Use of Eq.~\eqref{mean_field} type approximation along with the commutation rule and assumption of no direct transition between ground ($g$) and second excited state ($f$) provide Eq.~\eqref{eq:single_exc_coh} to Eq.~\eqref{eq:cross_coh}. The equations of motion for the coherences can be truncated to first order in their respective coupling constants; specifically, 
$\langle B_m^\dagger B_n \rangle \equiv W_{mn}$, 
$\langle D_m^\dagger D_n \rangle \equiv Z_{mn}$, and 
$\langle D_m^\dagger B_n \rangle \equiv R_{mn}$, 
with respect to $J$, $\mathcal{J}$, and $K$, respectively. This truncation allows us to obtain rates that are second order in the coupling parameters, leading to the following set of coherence equations.
\begin{equation}
\begin{aligned}
\label{eq:approximated_coherence}
\frac{\partial}{\partial t} W_{mn} 
&\approx i\frac{(E_m-E_n)}{\hbar}W_{mn}-\kappa W_{mn} + \frac{i}{\hbar} J_{nm} \left[ (\Pi_m - P_m) P_n - (\Pi_n - P_n) P_m \right] \\
\frac{\partial}{\partial t} Z_{mn} 
&\approx i\frac{(\epsilon_m-\epsilon_n)-(E_m-E_n)}{\hbar}Z_{mn}-(\kappa +r)Z_{mn} 
+ \frac{i}{\hbar} J_{nm} \left[ (P_m - N_m) N_n - N_m (P_n - N_n) \right] \\
\frac{\partial}{\partial t} R_{mn} 
&\approx i\,\frac{\left(\epsilon_m - E_m - E_n\right)}{\hbar}\,R_{mn}-(\kappa +r/2)R_{mn} 
+ \frac{i}{\hbar} K^{*}_{mn} \left[ (P_m - N_m) P_n - N_m (\Pi_n - P_n) \right]
\end{aligned}
\end{equation}
Now, considering rapid decay of equation of motion coherences compared to the population change we can eliminate the time dependence in Eq.~\eqref{eq:approximated_coherence} and followed by the substitution in population equations [Eq.~\eqref{eq:population_eqn} and Eq.\eqref{eq:pop_eqn_second}], we obtain the rate equation description Eq.\eqref{eq:rate_description} and Eq.\eqref{eq:rate_2nd_exc}.

\bibliography{main}

@article{engel2007evidence,
  title={Evidence for wavelike energy transfer through quantum coherence in photosynthetic systems},
  author={Engel, Gregory S and Calhoun, Tessa R and Read, Elizabeth L and Ahn, Tae-Kyu and Man{\v{c}}al, Tom{\'a}{\v{s}} and Cheng, Yuan-Chung and Blankenship, Robert E and Fleming, Graham R},
  journal={Nature},
  volume={446},
  number={7137},
  pages={782--786},
  year={2007},
  publisher={Nature Publishing Group UK London}
}

@article{bubilaitis2025compact,
  title={Compact modeling of highly excited linear aggregates using generalized quantum particles},
  author={Bubilaitis, Vytautas and Abramavicius, Darius},
  journal={Chemical Physics},
  volume={588},
  pages={112445},
  year={2025},
  publisher={Elsevier}
}

@article{rebentrost2009environment,
  title={Environment-assisted quantum transport},
  author={Rebentrost, Patrick and Mohseni, Masoud and Kassal, Ivan and Lloyd, Seth and Aspuru-Guzik, Al{\'a}n},
  journal={New Journal of Physics},
  volume={11},
  number={3},
  pages={033003},
  year={2009}
}

@article{calhoun2009quantum,
  title={Quantum coherence enabled determination of the energy landscape in light-harvesting complex II},
  author={Calhoun, Tessa R and Ginsberg, Naomi S and Schlau-Cohen, Gabriela S and Cheng, Yuan-Chung and Ballottari, Matteo and Bassi, Roberto and Fleming, Graham R},
  journal={The Journal of Physical Chemistry B},
  volume={113},
  number={51},
  pages={16291--16295},
  year={2009},
  publisher={ACS Publications}
}

@article{collini2010coherently,
  title={Coherently wired light-harvesting in photosynthetic marine algae at ambient temperature},
  author={Collini, Elisabetta and Wong, Cathy Y and Wilk, Krystyna E and Curmi, Paul MG and Brumer, Paul and Scholes, Gregory D},
  journal={Nature},
  volume={463},
  number={7281},
  pages={644--647},
  year={2010},
  publisher={Nature Publishing Group UK London}
}

@article{ishizaki2009theoretical,
  title={Theoretical examination of quantum coherence in a photosynthetic system at physiological temperature},
  author={Ishizaki, Akihito and Fleming, Graham R},
  journal={Proceedings of the National Academy of Sciences},
  volume={106},
  number={41},
  pages={17255--17260},
  year={2009},
  publisher={National Academy of Sciences}
}

@article{ishizaki2009unified,
  title={Unified treatment of quantum coherent and incoherent hopping dynamics in electronic energy transfer: Reduced hierarchy equation approach},
  author={Ishizaki, Akihito and Fleming, Graham R},
  journal={The Journal of chemical physics},
  volume={130},
  number={23},
  year={2009},
  publisher={AIP Publishing}
}

@article{dutta2025quantum,
  title={Quantum and classical effects in system-bath correlations and optical line shapes},
  author={Dutta, Rajesh and Reppert, Mike},
  journal={Physical Review A},
  volume={111},
  number={2},
  pages={022210},
  year={2025},
  publisher={APS}
}

@article{jansen2024electronically,
  title={Electronically excited states in cylindrical molecular aggregates: Exciton delocalization, dynamics, and optical response},
  author={Jansen, TLC and G{\"u}nther, Lisa M and Knoester, J and K{\"o}hler, J{\"u}rgen},
  journal={Chemical Physics Reviews},
  volume={5},
  number={4},
  year={2024},
  publisher={AIP Publishing}
}

@article{maly2018signatures,
  title = {Signatures of exciton delocalization and exciton--exciton annihilation in fluorescence-detected two-dimensional coherent spectroscopy},
  author = {Mal\'{y}, Pavel and Man\v{c}al, Tom\'{a}\v{s}},
  journal = {The Journal of Physical Chemistry Letters},
  volume = {9},
  number = {19},
  pages = {5654--5659},
  year = {2018},
  publisher = {ACS Publications}
}

@article{barford2024using,
  title={Using the Haken--Strobl--Reineker model to determine the temperature dependence of the diffusion coefficient},
  author={Barford, William},
  journal={Journal of chemical theory and computation},
  volume={20},
  number={15},
  pages={6510--6517},
  year={2024},
  publisher={ACS Publications}
}

@article{yuen2014coherent,
  title={Coherent exciton dynamics in supramolecular light-harvesting nanotubes revealed by ultrafast quantum process tomography},
  author={Yuen-Zhou, Joel and Arias, Dylan H and Eisele, Dorthe M and Steiner, Colby P and Krich, Jacob J and Bawendi, Moungi G and Nelson, Keith A and Aspuru-Guzik, Al{\'a}n},
  journal={ACS nano},
  volume={8},
  number={6},
  pages={5527--5534},
  year={2014},
  publisher={ACS Publications}
}

@article{kassal2013does,
  title={Does coherence enhance transport in photosynthesis?},
  author={Kassal, Ivan and Yuen-Zhou, Joel and Rahimi-Keshari, Saleh},
  journal={The journal of physical chemistry letters},
  volume={4},
  number={3},
  pages={362--367},
  year={2013},
  publisher={ACS Publications}
}

@article{valkunas2009exciton,
  title={Exciton migration and fluorescence quenching in LHCII aggregates: Target analysis using a simple nonlinear annihilation scheme},
  author={Valkunas, Leonas and van Stokkum, Ivo HM and Berera, Rudi and van Grondelle, Rienk},
  journal={Chemical Physics},
  volume={357},
  number={1-3},
  pages={17--20},
  year={2009},
  publisher={Elsevier}
}

@article{valkunas1995nonlinear,
  title={Nonlinear annihilation of excitations in photosynthetic systems},
  author={Valkunas, Leonas and Trinkunas, Gediminas and Liuolia, Vladas and Van Grondelle, R},
  journal={Biophysical journal},
  volume={69},
  number={3},
  pages={1117--1129},
  year={1995},
  publisher={Elsevier}
}

@article{bubilaitis2024signatures,
  title={Signatures of exciton--exciton annihilation in 2DES spectra including up to six-wave mixing processes},
  author={Bubilaitis, Vytautas and Abramavicius, Darius},
  journal={The Journal of Chemical Physics},
  volume={161},
  number={10},
  year={2024},
  publisher={AIP Publishing}
}

@article{knoester1995unusual,
  title={Unusual behavior of two-photon absorption from three-level molecules in a one-dimensional lattice},
  author={Knoester, Jasper and Spano, Frank C},
  journal={Physical review letters},
  volume={74},
  number={14},
  pages={2780},
  year={1995},
  publisher={APS}
}

@article{moix2011efficient,
  title={Efficient energy transfer in light-harvesting systems, III: The influence of the eighth bacteriochlorophyll on the dynamics and efficiency in FMO},
  author={Moix, Jeremy and Wu, Jianlan and Huo, Pengfei and Coker, David and Cao, Jianshu},
  journal={The Journal of Physical Chemistry Letters},
  volume={2},
  number={24},
  pages={3045--3052},
  year={2011},
  publisher={ACS Publications}
}

@article{dutta2019delocalization,
  title={Delocalization and quantum entanglement in physical systems},
  author={Dutta, Rajesh and Bagchi, Biman},
  journal={The Journal of Physical Chemistry Letters},
  volume={10},
  number={9},
  pages={2037--2043},
  year={2019},
  publisher={ACS Publications}
}

@article{collini2009coherent,
  title={Coherent intrachain energy migration in a conjugated polymer at room temperature},
  author={Collini, Elisabetta and Scholes, Gregory D},
  journal={science},
  volume={323},
  number={5912},
  pages={369--373},
  year={2009},
  publisher={American Association for the Advancement of Science}
}

@article{maly2020wavelike,
  title={From wavelike to sub-diffusive motion: exciton dynamics and interaction in squaraine copolymers of varying length},
  author={Mal{\`y}, Pavel and L{\"u}ttig, Julian and Turkin, Arthur and Dost{\'a}l, Jakub and Lambert, Christoph and Brixner, Tobias},
  journal={Chemical science},
  volume={11},
  number={2},
  pages={456--466},
  year={2020},
  publisher={Royal Society of Chemistry}
}

@article{barford2014theory,
  title={Theory of exciton transfer and diffusion in conjugated polymers},
  author={Barford, William and Tozer, Oliver Robert},
  journal={The Journal of chemical physics},
  volume={141},
  number={16},
  year={2014},
  publisher={AIP Publishing}
}

@article{bittner2014noise,
  title={Noise-induced quantum coherence drives photo-carrier generation dynamics at polymeric semiconductor heterojunctions},
  author={Bittner, Eric R and Silva, Carlos},
  journal={Nature communications},
  volume={5},
  number={1},
  pages={3119},
  year={2014},
  publisher={Nature Publishing Group UK London}
}

@article{cao2009optimization,
  title={Optimization of exciton trapping in energy transfer processes},
  author={Cao, Jianshu and Silbey, Robert J},
  journal={The Journal of Physical Chemistry A},
  volume={113},
  number={50},
  pages={13825--13838},
  year={2009},
  publisher={ACS Publications}
}

@article{wu2010efficient,
  title={Efficient energy transfer in light-harvesting systems, I: optimal temperature, reorganization energy and spatial--temporal correlations},
  author={Wu, Jianlan and Liu, Fan and Shen, Young and Cao, Jianshu and Silbey, Robert J},
  journal={New Journal of Physics},
  volume={12},
  number={10},
  pages={105012},
  year={2010},
  publisher={IOP Publishing}
}

@article{chen2011excitation,
  title={Excitation energy transfer in a non-Markovian dynamical disordered environment: Localization, narrowing, and transfer efficiency},
  author={Chen, Xin and Silbey, Robert J},
  journal={The Journal of Physical Chemistry B},
  volume={115},
  number={18},
  pages={5499--5509},
  year={2011},
  publisher={ACS Publications}
}

@article{dutta2021excitation,
  title={Excitation energy transfer efficiency in fluctuating environments: Role of quantum coherence in the presence of memory effects},
  author={Dutta, Rajesh and Bagchi, Biman},
  journal={The Journal of Physical Chemistry A},
  volume={125},
  number={22},
  pages={4695--4704},
  year={2021},
  publisher={ACS Publications}
}

@article{zhang2015delocalized,
  title={Delocalized quantum states enhance photocell efficiency},
  author={Zhang, Yiteng and Oh, Sangchul and Alharbi, Fahhad H and Engel, Gregory S and Kais, Sabre},
  journal={Physical chemistry chemical physics},
  volume={17},
  number={8},
  pages={5743--5750},
  year={2015},
  publisher={Royal Society of Chemistry}
}

@article{zhang2016dark,
  title={Dark states enhance the photocell power via phononic dissipation},
  author={Zhang, Yiteng and Wirthwein, Aaron and Alharbi, Fahhad H and Engel, Gregory S and Kais, Sabre},
  journal={Physical Chemistry Chemical Physics},
  volume={18},
  number={46},
  pages={31845--31849},
  year={2016},
  publisher={Royal Society of Chemistry}
}

@article{sarkar2020environment,
  title={Environment-assisted and environment-hampered efficiency at maximum power in a molecular photocell},
  author={Sarkar, Subhajit and Dubi, Yonatan},
  journal={The Journal of Physical Chemistry C},
  volume={124},
  number={28},
  pages={15115--15122},
  year={2020},
  publisher={ACS Publications}
}

@article{haken1972coupled,
  title={The coupled coherent and incoherent motion of excitons and its influence on the line shape of optical absorption},
  author={Haken, Hermann and Reineker, Peter},
  journal={Zeitschrift f{\"u}r Physik},
  volume={249},
  number={3},
  pages={253--268},
  year={1972},
  publisher={Springer}
}

@article{haken1973exactly,
  title={An exactly solvable model for coherent and incoherent exciton motion},
  author={Haken, Hermann and Strobl, Gert},
  journal={Zeitschrift f{\"u}r Physik A Hadrons and nuclei},
  volume={262},
  number={2},
  pages={135--148},
  year={1973},
  publisher={Springer}
}

@article{silbey1976electronic,
  title={Electronic energy transfer in molecular crystals},
  author={Silbey, Robert},
  journal={Annual Review of Physical Chemistry},
  volume={27},
  number={1},
  pages={203--223},
  year={1976},
  publisher={Annual Reviews 4139 El Camino Way, PO Box 10139, Palo Alto, CA 94303-0139, USA}}

@article{moix2013coherent,
  title   = {Coherent quantum transport in disordered systems: {I}. The influence of dephasing on the transport properties and absorption spectra in one-dimensional systems},
  author  = {Moix, Jeremy M. and Khasin, Michael and Cao, Jianshu},
  journal = {New Journal of Physics},
  volume  = {15},
  number  = {8},
  pages   = {085010},
  year    = {2013}
}

@article{lee2015coherent,
  title={Coherent quantum transport in disordered systems: A unified polaron treatment of hopping and band-like transport},
  author={Lee, Chee Kong and Moix, Jeremy and Cao, Jianshu},
  journal={The Journal of chemical physics},
  volume={142},
  number={16},
  year={2015},
  publisher={AIP Publishing}
}

@article{dutta2016effects,
  title={Effects of dynamic disorder on exciton migration: Quantum diffusion, coherences, and energy transfer},
  author={Dutta, Rajesh and Bagchi, Biman},
  journal={The Journal of Chemical Physics},
  volume={145},
  number={16},
  year={2016},
  publisher={AIP Publishing}
}

@article{chuang2016quantum,
  title={Quantum diffusion on molecular tubes: Universal scaling of the 1D to 2D transition},
  author={Chuang, Chern and Lee, Chee Kong and Moix, Jeremy M and Knoester, Jasper and Cao, Jianshu},
  journal={Physical review letters},
  volume={116},
  number={19},
  pages={196803},
  year={2016},
  publisher={APS}
}

@article{chuang2021universal,
  title={Universal scalings in two-dimensional anisotropic dipolar excitonic systems},
  author={Chuang, Chern and Cao, Jianshu},
  journal={Physical Review Letters},
  volume={127},
  number={4},
  pages={047402},
  year={2021},
  publisher={APS}
}

@article{karamlou2022quantum,
  title={Quantum transport and localization in 1d and 2d tight-binding lattices},
  author={Karamlou, Amir H and Braum{\"u}ller, Jochen and Yanay, Yariv and Di Paolo, Agustin and Harrington, Patrick M and Kannan, Bharath and Kim, David and Kjaergaard, Morten and Melville, Alexander and Muschinske, Sarah and others},
  journal={npj Quantum Information},
  volume={8},
  number={1},
  pages={35},
  year={2022},
  publisher={Nature Publishing Group UK London}
}

@article{blach2025environment,
  title={Environment-assisted quantum transport of excitons in perovskite nanocrystal superlattices},
  author={Blach, Daria D and Lumsargis-Roth, Victoria A and Chuang, Chern and Clark, Daniel E and Deng, Shibin and Williams, Olivia F and Li, Christina W and Cao, Jianshu and Huang, Libai},
  journal={Nature communications},
  volume={16},
  number={1},
  pages={1270},
  year={2025},
  publisher={Nature Publishing Group UK London}
}

@article{spano1989nonlinear,
  title={Nonlinear susceptibilities of molecular aggregates: Enhancement of $\chi$ (3) by size},
  author={Spano, Francis C and Mukamel, Shaul},
  journal={Physical Review A},
  volume={40},
  number={10},
  pages={5783},
  year={1989},
  publisher={APS}
}

@article{kuhn1996two,
  title={Two-exciton spectroscopy of photosynthetic antenna complexes: Collective oscillator analysis},
  author={K{\"u}hn, O and Chernyak, V and Mukamel, S},
  journal={The Journal of chemical physics},
  volume={105},
  number={19},
  pages={8586--8601},
  year={1996},
  publisher={American Institute of Physics}
}

@article{spano1991fermion,
  title={Fermion excited states in one-dimensional molecular aggregates with site disorder: nonlinear optical response},
  author={Spano, Frank C},
  journal={Physical review letters},
  volume={67},
  number={24},
  pages={3424},
  year={1991},
  publisher={APS}
}

@article{fidder1993observation,
  title={Observation of the one-exciton to two-exciton transition in a J aggregate},
  author={Fidder, Henk and Knoester, Jasper and Wiersma, Douwe A},
  journal={The Journal of chemical physics},
  volume={98},
  number={8},
  pages={6564--6566},
  year={1993},
  publisher={AMER INST PHYSICS}
}

@article{abramavicius2013exciton,
  title={Exciton-exciton scattering in a one-dimensional J aggregate},
  author={Abramavicius, D},
  journal={Europhysics Letters},
  volume={101},
  number={5},
  pages={57007},
  year={2013},
  publisher={IOP Publishing}
}

@article{barzda2001singlet,
  title={Singlet--singlet annihilation kinetics in aggregates and trimers of LHCII},
  author={Barzda, V and Gulbinas, V and Kananavicius, R and Cervinskas, V and Van Amerongen, H and Van Grondelle, R and Valkunas, L},
  journal={Biophysical journal},
  volume={80},
  number={5},
  pages={2409--2421},
  year={2001},
  publisher={Elsevier}
}

@article{bittner1994ultrafast,
  title={Ultrafast excitation energy transfer and exciton-exciton annihilation processes in isolated light harvesting complexes of photosystem II (LHC II) from spinach},
  author={Bittner, Thomas and Irrgang, Klaus-Dieter and Renger, Gernot and Wasielewski, Michael R},
  journal={The Journal of Physical Chemistry},
  volume={98},
  number={46},
  pages={11821--11826},
  year={1994},
  publisher={ACS Publications}
}

@article{dostal2018direct,
  title={Direct observation of exciton--exciton interactions},
  author={Dost{\'a}l, Jakub and Fennel, Franziska and Koch, Federico and Herbst, Stefanie and W{\"u}rthner, Frank and Brixner, Tobias},
  journal={Nature Communications},
  volume={9},
  number={1},
  pages={2466},
  year={2018},
  publisher={Nature Publishing Group UK London}
}

@article{kumar2023exciton,
  title={Exciton annihilation in molecular aggregates suppressed through quantum interference},
  author={Kumar, Sarath and Dunn, Ian S and Deng, Shibin and Zhu, Tong and Zhao, Qiuchen and Williams, Olivia F and Tempelaar, Roel and Huang, Libai},
  journal={Nature chemistry},
  volume={15},
  number={8},
  pages={1118--1126},
  year={2023},
  publisher={Nature Publishing Group UK London}
}

@Article{Deng2020,
author={Deng, Shibin
and Shi, Enzheng
and Yuan, Long
and Jin, Linrui
and Dou, Letian
and Huang, Libai},
title={Long-range exciton transport and slow annihilation in two-dimensional hybrid perovskites},
journal={Nature Communications},
year={2020},
month={Jan},
day={31},
volume={11},
number={1},
pages={664},
issn={2041-1723},
doi={10.1038/s41467-020-14403-z},
url={https://doi.org/10.1038/s41467-020-14403-z}
}

@article{Zanni2016ACSPhotonics,
author = {Ostrander, Joshua S. and Serrano, Arnaldo L. and Ghosh, Ayanjeet and Zanni, Martin T.},
title = {Spatially Resolved Two-Dimensional Infrared Spectroscopy via Wide-Field Microscopy},
journal = {ACS Photonics},
volume = {3},
number = {7},
pages = {1315-1323},
year = {2016},

}

@article{Libai2017ACR,
author = {Zhu, Tong and Wan, Yan and Huang, Libai},
title = {Direct Imaging of Frenkel Exciton Transport by Ultrafast Microscopy},
journal = {Accounts of Chemical Research},
volume = {50},
number = {7},
pages = {1725-1733},
year = {2017},
}

@article{Libai2013ARMR,
   author = "Huang, Libai and Cheng, Ji-Xin",
   title = "Nonlinear Optical Microscopy of Single Nanostructures", 
   journal= "Annual Review of Materials Research",
   year = "2013",
   volume = "43",
   number = "Volume 43, 2013",
   pages = "213-236",
   doi = "https://doi.org/10.1146/annurev-matsci-071312-121652",
   url = "https://www.annualreviews.org/content/journals/10.1146/annurev-matsci-071312-121652",
   publisher = "Annual Reviews",
   issn = "1545-4118",
   type = "Journal Article",
  }

@Article{Wan2015,
author={Wan, Yan
and Guo, Zhi
and Zhu, Tong
and Yan, Suxia
and Johnson, Justin
and Huang, Libai},
title={Cooperative singlet and triplet exciton transport in tetracene crystals visualized by ultrafast microscopy},
journal={Nature Chemistry},
year={2015},
month={Oct},
day={01},
volume={7},
number={10},
pages={785-792},
issn={1755-4349},
doi={10.1038/nchem.2348},
url={https://doi.org/10.1038/nchem.2348}
}

@article{sohoni2024optically,
  title={Optically accessible long-lived electronic biexcitons at room temperature in strongly coupled H-aggregates},
  author={Sohoni, Siddhartha and Ghosh, Indranil and Nash, Geoffrey T and Jones, Claire A and Lloyd, Lawson T and Li, Beiye C and Ji, Karen L and Wang, Zitong and Lin, Wenbin and Engel, Gregory S},
  journal={Nature Communications},
  volume={15},
  number={1},
  pages={8280},
  year={2024},
  publisher={Nature Publishing Group UK London}
}

@article{sundstrom1988annihilation,
  title={Annihilation of singlet excitons in J aggregates of pseudoisocyanine (PIC) studied by pico-and subpicosecond spectroscopy},
  author={Sundstr{\"o}m, V and Gillbro, T and Gadonas, RA and Piskarskas, A},
  journal={The Journal of chemical physics},
  volume={89},
  number={5},
  pages={2754--2762},
  year={1988},
  publisher={American Institute of Physics}
}

@article{van1995dynamics,
  title={The dynamics of one-dimensional excitons in liquids},
  author={van Burgel, Mirjam and Wiersma, Douwe A and Duppen, Koos},
  journal={The Journal of Chemical Physics},
  volume={102},
  number={1},
  pages={20--33},
  year={1995},
  publisher={AIP Publishing}
}

@article{gadonas1997wavelength,
  title={Wavelength and intensity-dependent transient degenerate four-wave mixing in pseudoisocyanine J-aggregates},
  author={Gadonas, R and Feller, K-H and Pugzlys, A and Jonusauskas, Gediminas and Oberl{\'e}, J and Rulliere, C},
  journal={The Journal of chemical physics},
  volume={106},
  number={20},
  pages={8374--8383},
  year={1997},
  publisher={American Institute of Physics}
}

@book{van2000photosynthetic,
  title={Photosynthetic excitons},
  author={Van Amerongen, Herbert and Van Grondelle, Rienk and others},
  year={2000},
  publisher={World Scientific}
}

@article{gaivzauskas1995annihilation,
  title={Annihilation enhanced four-wave mixing in molecular aggregates},
  author={Gai{\v{z}}auskas, E and Feller, K-H and Gadonas, R},
  journal={Optics communications},
  volume={118},
  number={3-4},
  pages={360--366},
  year={1995},
  publisher={Elsevier}
}

@article{malyshev1999exciton,
  title={Exciton--exciton annihilation in linear molecular aggregates at low temperature},
  author={Malyshev, VA and Glaeske, H and Feller, K-H},
  journal={Chemical physics letters},
  volume={305},
  number={1-2},
  pages={117--122},
  year={1999},
  publisher={Elsevier}
}

@article{ryzhov2001low,
  title={Low-temperature kinetics of exciton--exciton annihilation of weakly localized one-dimensional Frenkel excitons},
  author={Ryzhov, IV and Kozlov, GG and Malyshev, VA and Knoester, J},
  journal={The Journal of Chemical Physics},
  volume={114},
  number={12},
  pages={5322--5329},
  year={2001},
  publisher={American Institute of Physics}
}

@article{renger1997multiple,
  title={Multiple exciton effects in molecular aggregates: Application to a photosynthetic antenna complex},
  author={Renger, Thomas and May, Volkhard},
  journal={Physical review letters},
  volume={78},
  number={17},
  pages={3406},
  year={1997},
  publisher={APS}
}

@article{renger1997theory,
  title={Theory of multiple exciton effects in the photosynthetic antenna complex LHC-II},
  author={Renger, Thomas and May, Volkhard},
  journal={The Journal of Physical Chemistry B},
  volume={101},
  number={37},
  pages={7232--7240},
  year={1997},
  publisher={ACS Publications}
}

@article{bruggemann2001microscopic,
  title={Microscopic theory of exciton annihilation: Application to the LH2 antenna system},
  author={Br{\"u}ggemann, Ben and Herek, Jennifer L and Sundstr{\"o}m, Villy and Pullerits, Tonu and May, Volkhard},
  journal={The Journal of Physical Chemistry B},
  volume={105},
  number={46},
  pages={11391--11394},
  year={2001},
  publisher={ACS Publications}
}

@article{tempelaar2017exciton,
  title={Exciton--exciton annihilation is coherently suppressed in H-Aggregates, but not in J-aggregates},
  author={Tempelaar, Roel and Jansen, Thomas LC and Knoester, Jasper},
  journal={The journal of physical chemistry letters},
  volume={8},
  number={24},
  pages={6113--6117},
  year={2017},
  publisher={ACS Publications}
}

@article{may2014kinetic,
  title={Kinetic theory of exciton--exciton annihilation},
  author={May, Volkhard},
  journal={The Journal of Chemical Physics},
  volume={140},
  number={5},
  year={2014},
  publisher={AIP Publishing}
}

@article{hader2016identification,
  title={Identification of effective exciton--exciton annihilation in squaraine--squaraine copolymers},
  author={Hader, Kilian and May, Volkhard and Lambert, Christoph and Engel, Volker},
  journal={Physical Chemistry Chemical Physics},
  volume={18},
  number={19},
  pages={13368--13374},
  year={2016},
  publisher={Royal Society of Chemistry}
}

@article{hoki2011excitation,
  title={Excitation of biomolecules by coherent vs. incoherent light: Model rhodopsin photoisomerization},
  author={Hoki, Kunihito and Brumer, Paul},
  journal={Procedia Chemistry},
  volume={3},
  number={1},
  pages={122--131},
  year={2011},
  publisher={Elsevier}
}

@article{jankovic2020exact,
  title={Exact description of excitonic dynamics in molecular aggregates weakly driven by light},
  author={Jankovi{\'c}, Veljko and Man{\v{c}}al, Tom{\'a}{\v{s}}},
  journal={The Journal of Chemical Physics},
  volume={153},
  number={24},
  year={2020},
  publisher={AIP Publishing}
}

@article{yang2020steady,
  title={Steady-state analysis of light-harvesting energy transfer driven by incoherent light: From dimers to networks},
  author={Yang, Pei-Yun and Cao, Jianshu},
  journal={The journal of physical chemistry letters},
  volume={11},
  number={17},
  pages={7204--7211},
  year={2020},
  publisher={ACS Publications}
}

@article{jung2020energy,
  title={Energy transfer under natural incoherent light: Effects of asymmetry on efficiency},
  author={Jung, Kenneth A and Brumer, Paul},
  journal={The Journal of Chemical Physics},
  volume={153},
  number={11},
  year={2020},
  publisher={AIP Publishing}
}

@article{dutta2024memory,
  title={Memory effects in the efficiency control of energy transfer under incoherent light excitation in noisy environments},
  author={Dutta, Rajesh and Bagchi, Biman},
  journal={The Journal of Chemical Physics},
  volume={160},
  number={24},
  year={2024},
  publisher={AIP Publishing}
}

@article{tutunnikov2023coherent,
  title={Coherent spatial control of wave packet dynamics on quantum lattices},
  author={Tutunnikov, Ilia and Chuang, Chern and Cao, Jianshu},
  journal={The Journal of Physical Chemistry Letters},
  volume={14},
  number={51},
  pages={11632--11639},
  year={2023},
  publisher={ACS Publications}
}

@article{wang2018laser,
  title={Laser pulse induced multi-exciton dynamics in molecular systems},
  author={Wang, Luxia and May, Volkhard},
  journal={Journal of Physics B: Atomic, Molecular and Optical Physics},
  volume={51},
  number={6},
  pages={064002},
  year={2018},
  publisher={IOP Publishing}
}

\end{document}